\documentclass[numberedappendix]{emulateapj}
\usepackage{apjfonts}

\usepackage{subfigure}

\newcommand{\detg}{{\sqrt{-g}}}
\newcommand{\del}{{\partial}}

\newcommand{\msun}{{\rm M_{\odot}}}

\newcommand{\dF}{{^{^*}\!\!F}}
\newcommand{\alf}{Alfv\'en}

\newcommand{\bB}{{\bf B}}

\newcommand{\bva}{{\bf v}_{\rm A}}
\newcommand{\va}{{\bf v}_{\rm A}}
\newcommand{\kdv}{({\bf{k}} \cdot {\bf v}_{\rm A})}

\newcommand{\km}{{\rm\,km}}
\newcommand{\yr}{{\rm\,yr}}

\newcommand{\gm}{{\rm\,g}}
\newcommand{\cm}{{\rm\,cm}}
\newcommand{\erg}{{\rm\,erg}}
\newcommand{\ergps}{{\rm\,erg~s^{-1}}}

\newcommand{\sE}{{\mathcal{E}}}
\newcommand{\IEDEN}{{u_g}}

\shortauthors{McKinney, J.C.}
\shorttitle{Black Hole Jet Formation: Numerical Models}
\begin{document}
\submitted{June 16, 2005}

\journalinfo{Web link for High Resolution document in footnote}

\title{Jet Formation in Black Hole Accretion Systems II: Numerical Models}

\author{Jonathan C. McKinney$^{1}$}
\altaffiltext{1}{Institute for Theory and Computation,
  Harvard-Smithsonian Center for Astrophysics, 60 Garden Street, MS
  51, Cambridge, MA 02138, USA\\
  High Res. Figures:
  \url{http://rainman.astro.uiuc.edu/\textasciitilde
    jon/jet2.pdf}}
\email{jmckinney@cfa.harvard.edu}

\begin{abstract}

In a companion theory paper, we presented a unified model of jet
formation. We suggested that primarily two types of relativistic
jets form near accreting black holes: a potentially
ultrarelativistic Poynting-dominated jet and a Poynting-baryon jet.
We showed that, for the collapsar model, the neutrino-driven
enthalpy flux (classic fireball model) is probably dominated by the
Blandford-Znajek energy flux, which predicts a jet Lorentz factor of
$\Gamma\sim 100-1000$. We showed that radiatively inefficient AGN,
such as M87, are synchrotron-cooling limited to $\Gamma\sim 2-10$.
Radiatively efficient x-ray binaries, such as GRS1915+105, are
Compton-drag limited to $\Gamma \lesssim 2$, but the jet may be
destroyed by Compton drag. However, the Poynting-baryon jet is a
collimated outflow with $\Gamma \sim 1-3$. Here we present general
relativistic hydromagnetic simulations of black hole accretion with
pair creation used to simulate jet formation in GRBs, AGN, and x-ray
binaries.  Our collapsar model shows the development of a patchy
``magnetic fireball'' with typically $\Gamma\sim 100-1000$ and a
Gaussian structure. Temporal variability of the jet is dominated by
toroidal field instabilities for $\gtrsim 10^2$ gravitational radii.
A broader Poynting-baryon jet with $\Gamma\sim 1.5$ could contribute
to a supernova.

\end{abstract}


\maketitle

\section{Introduction}\label{introduction}

General relativistic magnetohydrodynamics (GRMHD) is the black hole
mass-invariant (nonradiative) physics commonly used to describe black
hole accretion systems.  Such systems often exhibit jets.  However,
the observed jet properties, such as collimation and speed, are not
uniform between systems.  In \citet{mckinney2005b}, our goal was to
determine the unifying, or minimum number of, pieces of physics that
would explain jet formation associated with gamma-ray bursts (GRBs),
active galactic nuclei (AGN), and x-ray binaries.  Similar ideas have
been explored by other authors \citep{gc02,ghis03,meier2003}.  The
goal of \citet{mckinney2005b} was to understand jet {\it formation}
and explain the origin of the energy, composition, collimation, and
Lorentz factor.

This paper implements some of the theoretical ideas of
\citet{mckinney2005b} into a GRMHD numerical model.  In particular, we
study the collapsar model with pair creation and an effective model
for Fick diffusion of neutrons that contaminate the jet with an
electron-proton plasma.  The generic model parameters also allow the
numerical model to be interpreted as a model for AGN systems such as
M87.  We also comment on the applicability of the model to x-ray
binary systems.

\S~\ref{modelsum} summarizes the proposed unified model to explain jet
formation in all black hole accretion systems, where more details are
provided in \citet{mckinney2005b}.

\S~\ref{results} presents the results of a fiducial numerical model
corresponding to the collapsar model and describes the jet structure
and formation.  The relevance of the presented model to AGN and x-ray
binaries is discussed.  The jet is found to have some piece-wise
self-similar features, and curve fits are given for use by other
modellers. Acceleration of the GRB jet is found to occur over a large
range in radii rather than occurring close to the black hole.  The
introductory estimates of the Lorentz factor in \citet{mckinney2005b}
are refined based upon these numerical models.  The jet characteristic
structure, such as the formation of a double transonic (transfast)
flow, is discussed.

\S~\ref{discussion} discusses the results and their possible
implications.  The results are compared to similar investigations, and
the limitations of the models presented are discussed.

\S~\ref{conclusions} summarizes the key points.

In appendix~\ref{GRMHD}, we give the GRMHD equations of motion solved
when accounting for pair creation.  See also \citet{mckinney2005b} for
relevant discussions.  These are the equations numerically solved that
give the results discussed in section~\ref{results}.  These are
straightforward and how one handles the details of the pair creation
ends up not making any qualitative (or much quantitative) difference
in the results, hence why the material is in the appendix.  That is,
the energetics of the Poynting-dominated jet are dominated by the
electromagnetic field.  Appendix~\ref{chars} summarizes the well-known
characteristics of the GRMHD equations and other surfaces of physical
interest used in the discussion in section~\ref{results}.

\section{GRMHD Pair Injection Model of Jet Formation}\label{modelsum}

In \citet{mckinney2005b}, we showed that one can identify a small
subset of physics that can explain the jet energy, composition,
collimation, and Lorentz factor for all black hole accretion systems.

One key idea of that paper is that the terminal Lorentz factor is
determined by the toroidal magnetic energy per unit pair mass density
energy near the location where pairs can escape to infinity (beyond
the so-called ``stagnation surface'').  For GRBs, neutron diffusion is
crucial to explain (and limit) the Lorentz factor and explain the
baryon-pollution or baryon-contamination problem.  For AGN and x-ray
binaries, since a negligible number of baryons cross the field lines,
pair-loading from radiative annihilation is crucial to determine the
Lorentz factor of the Poynting-dominated jet since this determines the
rest-mass flux and density in the jet.

\begin{figure}
\includegraphics[width=3.3in,clip]{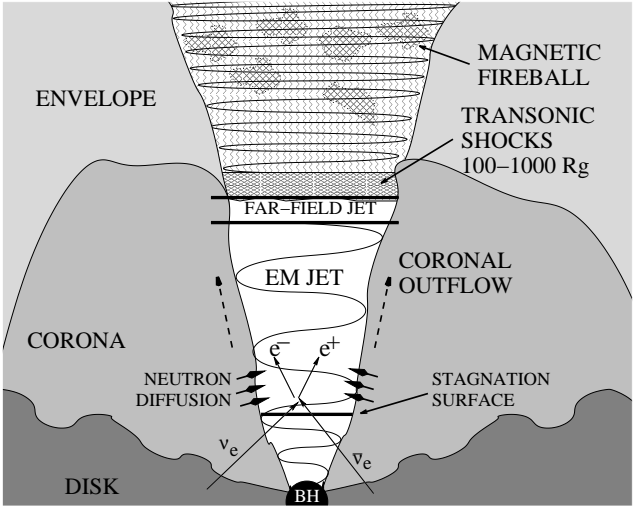}

\caption{Schematic of pair-production model and subsequent magnetic
fireball formation for GRB disks. Fireball is extremely optically
thick. Below a stagnation surface, pairs are accreted by the black
hole and so do not load the jet. Here $Rg=GM/c^2$. }
\label{model}
\end{figure}

\begin{figure}
\includegraphics[width=3.3in,clip]{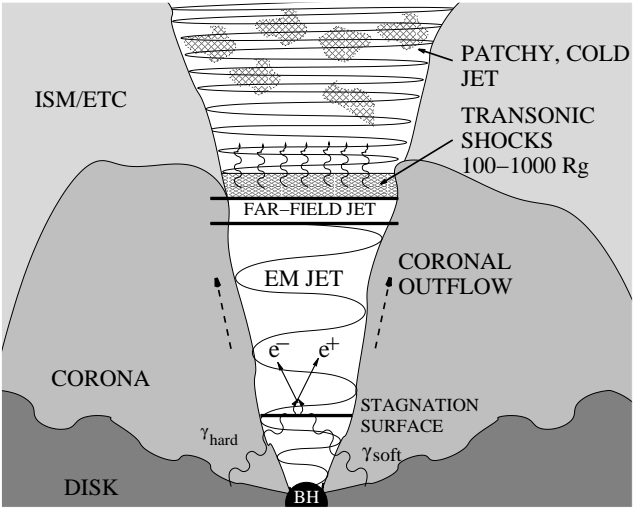}

\caption{Schematic of pair-production model and subsequent
shock-heating and emission. AGN jet is optically thin and emits
nonthermal and thermal synchrotron, while x-ray binary jet can be
marginally optically thick and emit via self-absorbed synchrotron and
by severe Compton drag.  Severe Compton drag can lead to destruction
of the Poynting-dominated jet. }
\label{modelagn}
\end{figure}

Figure~\ref{model} shows the basic picture for GRB systems, while
figure~\ref{modelagn} shows the basic picture for AGN and x-ray binary
systems.  An accreting, spinning black hole creates a magnetically
dominated funnel region around the polar axis.  The rotating black
hole drives a Poynting flux into the funnel region, where the Poynting
flux is associated with the coiling of poloidal magnetic field lines
into toroidal magnetic field lines.  The ideal MHD approximation holds
very well, and so only neutral particles such as photons, neutrinos,
and neutrons can cross the field lines and load the Poynting-dominated
jet that emerges.

The accretion disk emits neutrinos in a GRB model ($\gamma$-ray and
many soft photons for AGN and x-ray binaries) that annihilate and
pair-load the funnel region within some ``injection region.''  For GRB
systems, neutrons Fick-diffuse across the field lines and
collisionally decay into an electron-proton plasma.

Many pairs (any type) are swallowed by the black hole, but some escape
if beyond some ``stagnation surface,'' where the time-averaged
poloidal velocity is zero and positive beyond.  Pairs beyond the
stagnation surface are then accelerated by the Poynting flux in a
self-consistently generated collimated outflow.  In the
electromagnetic (EM) jet, the acceleration process corresponds to a
gradual uncoiling of the magnetic field and a release of the stored
magnetic energy that originated from the spin energy of the black
hole.

One key result of this paper is that the release of magnetic energy
need not be gradual once the toroidal field dominates the poloidal
field, in which case pinch (and perhaps kink) instabilities can occur
and lead to a nonlinear coupling (e.g. a shock) that converts Poynting
flux into enthalpy flux \citep{e93,begelman1998}.  In the proposed GRB
model, this conversion reaches equipartition and the jet becomes a
``magnetic fireball,'' where the toroidal field instabilities drive
large variations in the jet Lorentz factor and jet luminosity.

In \citet{mckinney2005b}, we showed that in AGN systems, nonthermal
synchrotron from shock-accelerated electrons and some thermal
synchrotron emission releases the shock energy until the synchrotron
cooling times are longer than the jet propagation time.  For AGN, jet
acceleration is negligible beyond the {\it extended} shock zone, as
suggested for blazars beyond the ``blazar zone''
\citep{sikora2005}. In x-ray binary systems, the shock is not as hot
and also unlike in the AGN (at least those like M87) case the jet can
be optically thick.  Thus these x-ray binary systems self-absorbed
synchrotron emit if they survive Compton drag.

For all these systems, at large radii patches of energy flux and
variations in the Lorentz factor develop due to toroidal
instabilities.  These patches in the jet could drive internal shocks
and at large radii they drive external shocks with the surrounding
medium.  The EM jet is also surrounded by a mildly relativistic matter
coronal outflow/jet/wind, which is a material extension of the corona
surrounding the disk.  This Poynting-baryon, coronal outflow
collimates the outer edge of the Poynting-dominated jet, which
otherwise internally collimates by hoop stresses.  The luminosity of
the Poynting-baryon jet is determined, like the Poynting-dominated
jet, by the mass accretion rate, disk thickness, and black hole spin.

In this paper, this unified model is studied numerically using
axisymmetric, nonradiative, GRMHD simulations to study the
self-consistent process of jet formation from black hole accretion
systems.  These simulations extend the work of \citet{mg04} by
including pair creation (and an effective neutron diffusion for
GRB-type systems) to self-consistently treat the creation of jet
matter, investigating a larger dynamic range in radius, and presenting
a more detailed analysis of the Poynting-dominated jet structure.

\subsection{Units and Notation}

The units in this paper have $G M = c = 1$, which sets the scale of
length ($r_g\equiv GM/c^2$) and time ($t_g\equiv GM/c^3$).  The mass
scale is determined by setting the (model-dependent) observed (or
inferred for GRB-type systems) mass accretion rate
($\dot{M}_0$[$g~s^{-1}$]) equal to the accretion rate through the
black hole horizon as measured in a simulation.  So the mass is scaled
by the mass accretion rate ($\dot{M}_0$) at the horizon ($r=r_H\equiv
r_L(1+\sqrt{1-j^2})$), such that $\rho_{0,disk}\equiv \dot{M}_0[r=r_H]
t_g/r_g^3$ and the mass scale is then just $m\equiv \rho_{0,disk}
r_g^3 = \dot{M}_0[r=r_H] t_g$.  Unless explicitly stated, the magnetic
field strength is given in Heaviside-Lorentz units, where the Gaussian
unit value is obtained by multiplying the Heaviside-Lorentz value by
$\sqrt{4\pi}$.

The value of $\rho_{0,disk}$ can be determined for different systems.
For example, a collapsar model with $\dot{M}=0.1\msun s^{-1}$ and
$M\approx 3\msun$, then $\rho_{0,disk}\approx 3.4\times 10^{10}{\rm
g}\cm^{-3}$.  M87 has a mass accretion rate of $\dot{M}\sim
10^{-2}\msun\yr^{-1}$ and a black hole mass of $M\approx 3\times
10^9\msun$ \citep{ho99,reynolds96} giving $\rho_{0,disk}\sim 10^{-16}
{\rm g}\cm^{-3}$.  GRS 1915+105 has a mass accretion rate of
$\dot{M}\sim 7\times 10^{-7}\msun\yr^{-1}$ \citep{mr94,mr99,fb04} with
a mass of $M\sim 14\msun$ \citep{greiner2001a}, but see
\citet{kaiser04}.  This gives $\rho_{0,disk}\sim 3\times 10^{-4}{\rm
g}\cm^{-3}$.  This disk density scales many of the results of the
paper.  The disk height ($H$) to radius ($R$) ratio is written as
$H/R$.

The GRMHD notation follows MTW.  For example, the 4-velocity
components are $u^\mu$ (contravariant) or $u_\mu$ (covariant).  For a
black hole with angular momentum $J=j GM^2/c$, $j=a/M$ is the
dimensionless Kerr parameter with $-1\le j\le 1$.  The rest-mass
density is given as $\rho_0$, internal energy density as $u$, magnetic
field 3-vector as $B^i\equiv \dF^{it}$, where $\dF$ is the dual of the
Faraday tensor.  The contravariant metric components are $g^{\mu\nu}$
and covariant components are $g_{\mu\nu}$, where beyond the
frame-dragging of the black hole the metric in Boyer-Lindquist
coordinates is approximately diagonal such that an orthonormal
contravariant component is $u^{\hat{\mu}}=\sqrt{g_{\mu\mu}} u^\mu$ for
any spatial component of $u^\mu$.  The comoving energy density is
$b^2/2$, where $b^\mu$ is the comoving magnetic field.  See
\citet{mg04} for details.

The Lorentz factor of the jet can be measured either as the current
time-dependent value, or, using information about the GRMHD system of
equations, one can estimate the Lorentz factor at large radii from
fluid quantities at small radii.  The Lorentz factor as measured by a
static observer at infinity is
\begin{equation}
\Gamma\equiv u^{\hat{t}}=u^t\sqrt{-g_{tt}}
\end{equation}
in Boyer-Lindquist coordinates, where no static observers exist inside
the ergosphere.  This is as opposed to $W\equiv u^t\sqrt{-1/g^{tt}}$,
which is the Lorentz factor as measured by the normal observer as used
by most numerical relativists.

In \citet{mckinney2005b} we show that the Lorentz factor and
$\phi$-velocity at large distances are
\begin{eqnarray}
\Gamma_{\infty}         & = & E = -h u_t    + \Phi\Omega_F B_\phi \\
u^{\hat{\phi}}_{\infty} & = & L = h u_\phi + \Phi B_\phi ,
\end{eqnarray}
where $h=(\rho_0+\IEDEN+p)/\rho_0$ is the specific enthalpy, $\Phi$ is
the conserved magnetic flux per unit rest-mass flux, $\Omega_F$ is the
conserved field rotation frequency, $B_\phi$ is the covariant toroidal
magnetic field, and $u^{\hat{\phi}}_{\infty}=u_{\phi,\infty}$.  Here
$E$ and $L$ simply represent the conserved energy and angular momentum
flux per unit rest-mass flux.  Notice the matter and electromagnetic
pieces are separable, such that
\begin{eqnarray}
\Gamma_{\infty} & = & \Gamma^{(MA)}_\infty + \Gamma^{(EM)}_\infty  \\
u_{\phi,\infty} & = & u^{(MA)}_{\phi,\infty} + u^{(MA)}_{\phi,\infty},
\end{eqnarray}
See \citet{mckinney2005b} for more details.

\section{Numerical Experiments}\label{results}

This section presents the results of GRMHD numerical models with pair
creation to self-consistently describe the Poynting-dominated and
Poynting-baryon jet formation process.  Our numerical scheme is HARM
\citep{gmt03}, a conservative, shock-capturing scheme for evolving the
equations of general relativistic MHD. Compared to the original HARM,
the inversion of conserved quantities to primitive variables is
performed by using a new faster and more robust two-dimensional
non-linear solver \citep{noble05}.  The new HARM also uses a parabolic
interpolation scheme \citep{colella84} rather than a linear
interpolation scheme.  The new HARM also uses an optimal TVD third
order Runge-Kutta time stepping \citep{shu97} rather than the
mid-point method.  For the problems under consideration, the parabolic
interpolation and third order time stepping method reduce the
truncation error significantly, including magnetically dominated
regions with $b^2/\rho_0\gg 1$.

Notice that no explicit reconnection model is included.  However, HARM
checks the effective resistivity by measuring the rest-mass flux
across field lines.  For the boundary between the coronal wind and the
Poynting-dominated jet, the total mass flux across the field line is
negligible compared to the rest-mass flux along the field line and the
pair creation rate.  An unresolved model would load field lines with
rest-mass that crosses field lines and not properly represent any
physical resistivity \citep{mckinney2004}.

\subsection{Computational Domain}

The computational domain is {\it axisymmetric}, with a grid that
typically extends from $r_{in} = 0.98 r_H$, where $r_H$ is the
horizon, to $r_{out} = 10^4r_g$, and from $\theta = 0$ to $\theta =
\pi$.  Full 3D calculations with this dynamic range are not possible
with today's computers.  Our numerical integrations are carried out on
a uniform grid in so-called ``modified KS'' (MKS) coordinates: $x_0,
x_1, x_2, x_3$, where $x_0 = t[{\rm KS}]$, $x_3 = \phi[{\rm KS}]$.
The radial coordinate is chosen to be
\begin{equation}\label{radius}
r = R_0 + e^{x^{n_r}_1} ,
\end{equation}
where $R_0$ is chosen to concentrate the grid zones toward the event
horizon (as $R_0$ is increased from $0$ to $r_H$) and $n_r$ is
controls the enhancement of inner to outer radial regions.  For
studies where the disk and jet interaction is of primary interest,
$R_0=0$ and $n_r=1$ are chosen.  For studies where in addition the
far-field jet is of interest, $R_0=-3$ and $n_r=10$ are chosen. The
$\theta$ coordinate is chosen to be
\begin{equation}\label{theta}
\theta = \pi x_2 + \frac{1}{2} (1 - h(r)) \sin(2\pi x_2) ,
\end{equation}
where $h(r)$ is used to concentrate grid zones toward the equator (as
$h$ is decreased from $1$ to $0$) or pole (as $h$ is increased from
$1$ to $2$).  The jet at large radii is resolved together with the
disk at small radii using
\begin{equation}\label{hofr}
h(r) = 2 - Q_j (r/r_{0j})^{-n_j g_j}
\end{equation}
with the parameters of $Q_j=1.3 - 1.8$, $r_{0j}=2.8$, $n_j=0.3$,
$r_{1j}=20$, $r_{2j}=80$, and
$g_j=g_j(r)=\frac{1}{2}+\frac{1}{\pi}atan(\frac{r-r_{2j}}{r_{1j}})$ is
used to control the transition between inner and outer radial regions.
To perform convergence tests, $Q_j$ and the overall radial and
$\theta$ resolution are varied.  An alternative to this fixed
refinement of the jet and disk is an adaptive refinement (see, e.g.,
\citealt{zm05}).

\subsection{Initial Conditions and Problem Setup}

All the experiments evolve a weakly magnetized torus around a Kerr
black hole in axisymmetry.  The focus of our numerical investigation
is to study a high resolution model of the collapsar model.  Other
simulations were performed, and the fiducial model's relevance to AGN
and x-ray binary systems is discussed at the end of the section.  A
generic model is used so the results mostly can be applied to any
black hole system.  A black hole spin of $j=0.9375$ is chosen, but
this produces similar results to models with $0.5\lesssim j\lesssim
0.97$, which includes most black hole accretion systems.

The initial conditions consist of an equilibrium torus which is a
``donut'' of plasma with a black hole at the center
\citep{fm76,ajs78}.  The donut is supported against gravity by
centrifugal and pressure forces. The solutions of \cite{fm76}
corresponding to $u^t u_\phi = const.$ are used. The initial inner
edge of the torus is set at $r_{edge} = 6$. The equation of state is a
gamma-law ideal gas with $\gamma = 4/3$, but other $\gamma$ lead to
similar results \citep{mg04}.  This donut is a solution to the
axisymmetric stationary equations, but the donut is unstable to global
nonaxisymmetric modes \citep{pp83}.  However, when the donut is
embedded in a weak magnetic field, the magnetorotational instability
(MRI) dominates those hydrodynamic modes. Small perturbations are
introduced in the velocity field to seed the instability.  The models
have $u^t u_\phi = 4.281$, the pressure maximum is located at $r_{max}
= 12$, the inner edge at $(r,\theta)=(6,\pi/2)$, and the outer edge at
$(r,\theta) = (42, \pi/2)$.

This is consistent with the accretion disk that is expected to form in
the collapsar model \citep{mw99}.  A GRB in the collapsar model is
formed after the collapse of a massive star as a black hole with
$M_{BH}\sim 3M_{\odot}$ and spin $j\sim 0.75-0.94$ accretes at
$\dot{M}_0\sim 0.1M_{\odot}/s$ from a torus of matter within
$r<200\km$ that has a height ($H$) to radius ($R$) ratio
(``thickness'') of $H/R\sim 0.2-0.4$.  \citet{mw99} find that an
energetic outflow develops in the polar region where $\rho\sim 10^{7}
\gm\cm^{-3}$ near the horizon, $\rho\sim 10^{5} \gm\cm^{-3}$ at $r\sim
200\km$, and $\rho\sim 10^{3-4} \gm\cm^{-3}$ at $r\sim 2000\km$ (see
model 14A in \citealt{mw99}).  In the parlance of collapsar models,
our $u^t u_\phi$ model corresponds to a $j_{16}\equiv j/(10^{16} \cm^2
{\rm s}^{-1})$ of $j_{16}\sim 5$, within the range suggested by
\citet{mw99} to produce a GRB.  Also, the $H/R\sim 0.2-0.4$ as
suggested that forms according to neutrino emission models
\citep{kohri2002,kohri2005}.

For radiatively inefficient AGN and x-ray binaries this is simply a
reasonable approximate model for the inner-radial accretion flow.
While a slightly thicker disk ($H/R\sim 0.9$) may be more appropriate
for radiatively inefficient flows, this is not expected to
significantly affect these results.  A study of the effect of disk
thickness, especially thin disks, is left for future work.

The orbital period at the pressure maximum $2 \pi (a +
\left(r_{max}/r_g\right)^{3/2}) t_g \simeq 267 t_g$, as measured by an
observer at infinity.  The model is run for $\Delta t = 1.4\times
10^4t_g$, which is about $52$ orbital periods at the pressure maximum
and about $1150$ orbital periods at the black hole horizon.  For the
collapsar model this is only $\sim 0.2$ seconds, and at the spin
chosen would correspond to the late phase of accretion when the
Poynting flux reaches its largest magnitude.  For the AGN M87 this
corresponds to $\sim 7~{\rm years}$.  For the x-ray binary GRS
1915+105 this corresponds to $\sim 1~{\rm second}$.

Notice that since the model is axisymmetric, disk turbulence is not
sustained after about $t\sim 3000t_g$ ; that is, the anti-dynamo
theorem prevails \citep{cowling34}.  However, while this affects the
disk accretion, this does not affect the evolution of the
Poynting-dominated jet.  That is, from the time of turbulent accretion
to ``laminar'' accretion, the funnel region is mostly unchanged.
Indeed, the far-field jet that has already formed is causally
disconnected from the region where the accretion disk would still be
turbulent.

For the collapsar model, at the quasi-steady state mass accretion rate
of $0.1 M_{\odot}/s$ the disk will last for $t\lesssim 0.42s$.  Thus,
this model requires a more extended disk or a fresh supply of plasma
into the disk from the surrounding stellar envelope through an
``accretion shock'' to generate a long duration GRBs
\citep{mw99}. However, since any such system modelled is in
quasi-steady state, the results here do not strongly depend on the
mass of the disk as long as the disk is not too massive, which would
require taking into account the self-gravity of the disk.

The numerical resolution of these models is $512\times 256$ compared
to $456^2$ in \citet{mg04}.  However, due to the enhanced $\theta$
grid, the resolution in the far-field jet region is $\sim 10$ times
larger.  Also, with the use of a parabolic interpolation scheme, the
overall resolution is additionally enhanced.  Compared to our previous
model this gives us an effective $\theta$ resolution of $\approx
9000$.

As with \citet{mg04}, into the initial torus is put a purely poloidal
magnetic field. The field can be described using a vector potential
with a single nonzero component $A_\phi \propto {\rm
MAX}(\rho_0/\rho_{0,max} - 0.2, 0)$ The field is therefore restricted
to regions with $\rho_0/\rho_{0,max} > 0.2$.  The field is normalized
so that the minimum ratio of gas to magnetic pressure is $100$.  The
equilibrium is therefore only weakly perturbed by the magnetic field.
It is, however, no longer stable \citep{bh91,gam04}.
\citet{hirose04,mg04} show that no initial large-scale net vertical
field is necessary, since a large-scale poloidal field is
self-consistently generated. The field connects the black hole horizon
(as observed by a physical observer) to large distances.  Apart from
the existence of an overall poloidal sign, the results are insensitive
to the details of the initial magnetic field geometry for any
physically motivated geometry \citep{mg04}.

\subsection{Floor vs. Pair Creation Model}

Numerical models often ``model'' the injection physics in the
Poynting-dominated jet by employing a so-called floor model.  This
model forces a minimum on the rest-mass density ($\rho_{fl}$) and
internal energy density ($u_{fl}$), which are usually set to several
orders of magnitude lower than the disk density.  Actually, however,
floor models have always been treated as a purely numerical invention
in order to avoid numerical artifacts associated with the inability to
numerically solve the equations of motion when the density is low.
Indeed, the idea was one should convergence test by gradually lowering
$\rho_{fl}$ and $u_{fl}$.  This is perhaps a reasonable for numerical
study of stars, but accretion flows are so hot (baryons with $T\gtrsim
10^{10}K$ in the thick disk state near the black hole) that pairs are
created when neutrinos annihilate (or when photons annihilate for
x-ray binary and AGN systems).

In previous numerical models of Poynting jets, the funnel region is
always completely evacuated, and floor-models necessarily generate
mass at least where the poloidal velocity $u^p=0$ at the stagnation
surface.  That is, matter inside the stagnation surface falls into the
black hole, while matter outside it is ejected as part of the jet.
Indeed, arbitrary floor-models violate the ideal-MHD condition far
away from the black hole where no pairs should be produced.  For
example, if a numerical model uses $\rho_{fl}\sim{\rm Const.}$, then
this leads to a completely unphysical model of pair creation and the
ideal-MHD approximation is violated for the entire length of the jet
in the funnel.

While pair creation has so far been ignored by those doing numerical
models of jets from black hole accretion systems, it has been shown
that the properties of the Poynting-dominated jet are almost
completely determined by the pair creation physics, which thus cannot
be ignored \citep{phi83,punsly1991,lev05}.

Our collapsar model accounts for pair creation and Fick diffusion of
neutrons.  In this paper, there is an injected rest-mass, enthalpy,
and momentum. For the fiducial collapsar-like model, rest-mass,
enthalpy, and momentum are injected with energy at infinity fractions
of, respectively, $f_\rho=0.05$, $f_h=0.45$, and $f_m=0.5$ as
described in \citet{mckinney2005b}.  The total energy at infinity
injected is determined by neutrino annihilation rates and Fick
diffusion rates.  The energy injection rate from neutrino annihilation
is based upon viscous disk models \citep{pwf99,mw99}, where the rate
of injection is determined by choosing a viscosity coefficient
$\alpha=0.01$ and mass accretion rate $\dot{M}_0=0.1\msun s^{-1}$ for
the collapsar model.  As discussed in that paper, the rest-mass is
dominated by electron-positron pairs only very close to the black
hole.

The value of $\alpha$ and $\dot{M}_0$ determine the energy injected in
proportion to the disk density per unit light crossing time ($t_g$).
Thus any system with comparable normalized energy injection rate would
follow similar results.  For example, the M87 model is similar, where
there the electron-positron pairs do not annihilate and represent the
true density in the jet.  In the M87 model there is no Fick diffusion,
yet the dimensionless model parameters are approximately the same.

The momentum direction is chosen by setting $u^\phi_{inj}=u^\phi$ and
$u^\theta_{inj}=0$, so that $f_m$ determines ${u^r}_{inj}$.  It turns
out that the choice of $f_h$, $f_m$, and the injected momentum
direction very weakly determine the jet structure for $r\gtrsim 6r_g$.
This is because a large fraction of the pairs are lost to the black
hole and magnetic forces, rather than the injected momentum, dominate
the motion of the plasma near the black hole.

Beyond this injection region, the rest-mass is dominated by
electron-protons created in collision events from neutrons that Fick
diffuse across the field lines into the jet region.  For the collapsar
model this coincidentally corresponds to the injected rest-mass
injected in this model.  Since the electron-positron pair annihilation
only gives an additional $\sim 10\%$ of internal energy, then pair
annihilation need not be considered, and the rest-mass injected
well-models the rest-mass injected as an electron-proton plasma from
Fick-diffused neutrons that collisionally decay.

This physical model of the injected particles is augmented when the
rest-mass or internal energy density reaches very low values.  If the
density drops below some ``floor'' value, then the density is returned
to the floor value.  The floor model chosen has $\rho_{fl}=10^{-7}
r^{-2.7}\rho_{0,disk}$ and $u_{fl}=10^{-9} r^{-2.7}\rho_{0,disk}$.
HARM keeps track of how often the density goes below the floors and
how this modifies the conservation of mass, energy, and angular
momentum.  The floor model contribution is negligible compared to the
physical injection model at all spatial locations and at all times.
The coefficient of the floor was chosen to be close to the resulting
density near the black hole horizon.  It was not chosen arbitrarily
small in order to avoid numerical difficulties in integrating the
equations when, rarely, the floor is activated in regions of large
magnetic energy density per unit rest-mass density.

\subsection{Boundary Conditions}

Two different models are chosen for the outer boundary condition.  One
model is called an ``outflow'' boundary condition, for which all
primitive variables are projected into the ghost zones while
forbidding inflow.  The other model is to inject matter at some
specified rate at the outer boundary, unless there is outflow.  This
would correspond to a Bondi-like infall for AGN and x-ray binaries, or
would correspond to the collapsing envelope of a massive star for
collapsar models.  In particular, unless there is outflow, the outer
boundary is set to inject mass at the free-fall rate, with a density
of $\rho_0 = 10^{-4} r^{-3/2}\rho_{0,disk}$ and $u = 10^{-6}
r^{-5/2}\rho_{0,disk}$ and no angular momentum.  Presupernova models
suggest that the infalling matter is at about $30\%$ the free-fall
rate we have chosen above \citep{mw99}, but this difference is
unlikely to significantly affect the jet formation.  Also, for their
collapsar model, the density structure in the equatorial plane varies
between $\rho_0\propto r^{-3/2}$ to $\rho_0\propto r^{-2}$ and the
internal energy is $u\propto r^{-5/2}$ to $u\propto r^{-2.7}$
\citep{mw99}.  This is similar to our simplified model, which happens
to also model a Bondi accretion for AGN and x-ray binaries.  Thus the
results are indicative of GRBs, AGN, and x-ray binaries.  The outer
grid radius corresponds to about $20$ presupernova core radii
\citep{ww95} or $\sim 10^{10}\km$ or about $1/10$th the entire star's
radius.

\subsection{Fiducial Model}\label{fiducial}

The overall character of the accretion flow is unchanged compared to
the descriptions given in \citet{mg04}.  The disk enters a long,
quasi-steady phase in which the accretion rates of rest-mass, angular
momentum, and energy onto the black hole fluctuate around a
well-defined mean.  Meanwhile, as in \citet{mg04}, a Poynting-dominated
jet and Poynting-baryon jet (coronal outflow) have formed.  The
Poynting-dominated jet forms once the ram pressure of the funnel material
is lower than the toroidal magnetic pressure.  Afterwards the baryons
are spread apart in the launch of a magnetic tube filled with pairs.
This occurs within $t\lesssim 500t_g$.

\begin{figure}
\subfigure{\includegraphics[width=3.3in,clip]{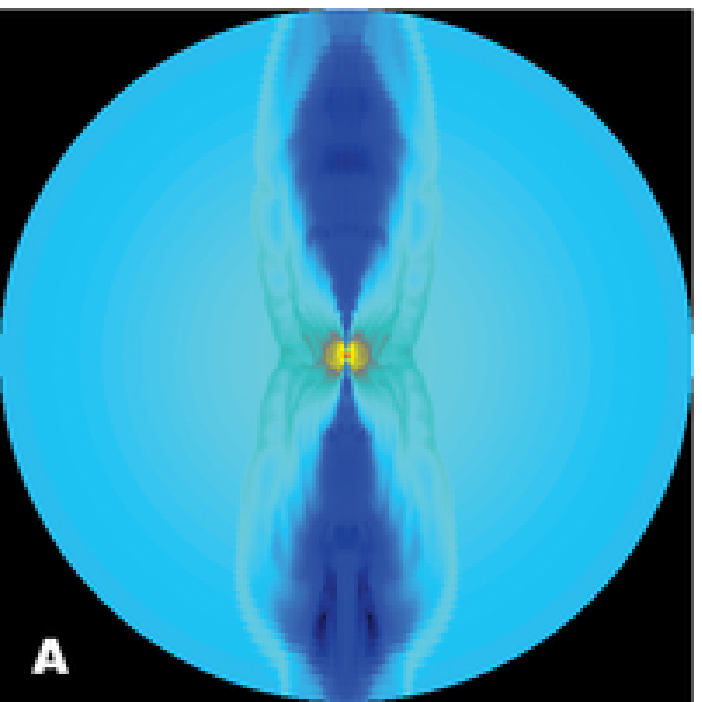}}
\subfigure{\includegraphics[width=3.3in,clip]{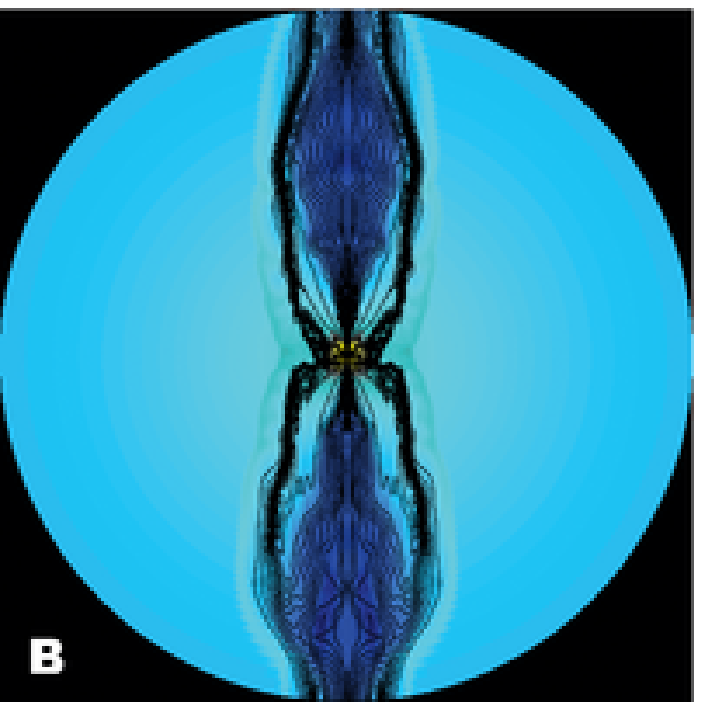}}

\caption{Jet has pummelled its way through presupernova core and
through $1/10$th of entire star.  Time is $t=1.4\times 10^4t_g$.
Panel (A) shows final distribution of $\log{\rho_0}$ on the Cartesian
plane. Black hole is located at center. Red is highest density and
black is lowest. Panel (B) shows magnetic field overlayed on top of
log of density. Outer scale is $r=10^4 GM/c^2$.}
\label{density}
\end{figure}

\begin{figure}
\subfigure{\includegraphics[width=3.3in,clip]{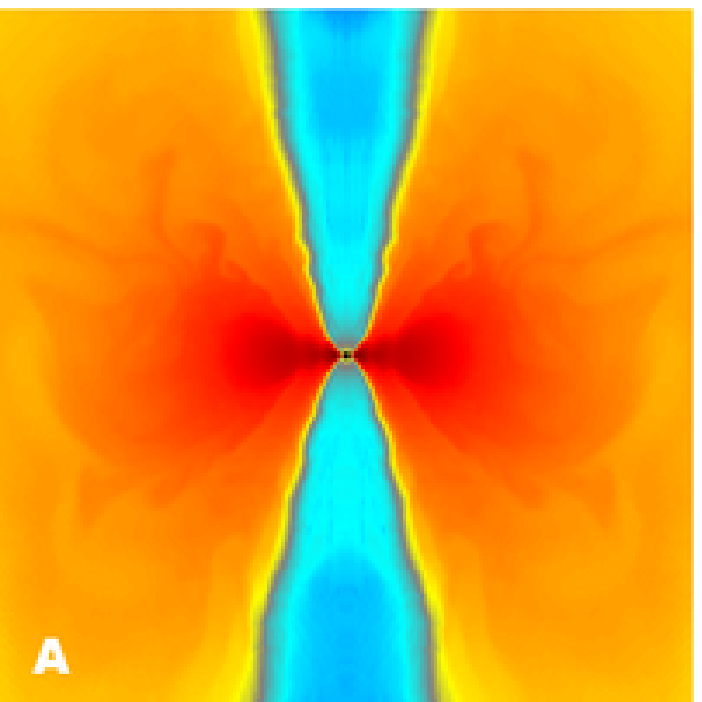}}
\subfigure{\includegraphics[width=3.3in,clip]{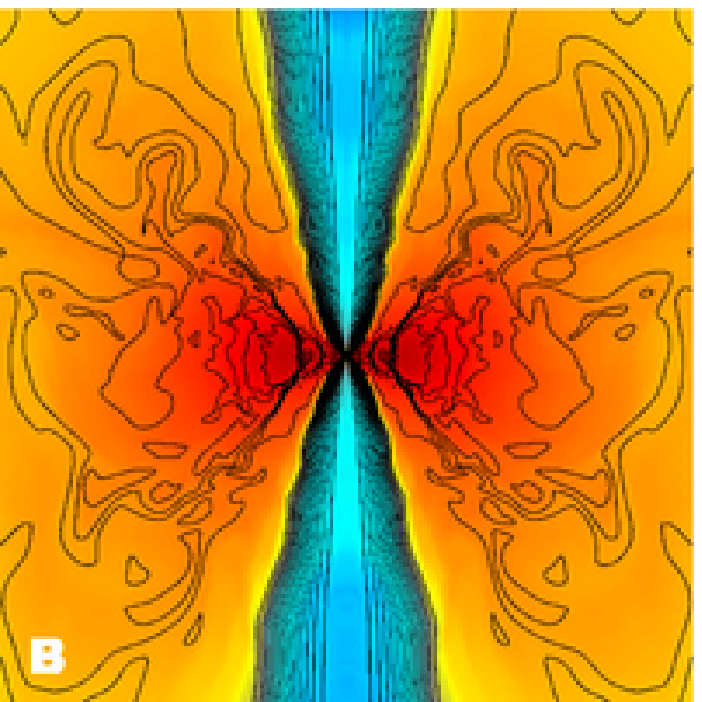}}

\caption{Strongest magnetic field near black hole in X-configuration
due to Blandford-Znajek effect and collimation of disk+coronal
outflow.  As in figure~\ref{density}, but outer scale is $r=10^2
GM/c^2$.  Time is $t=1.4\times 10^4t_g$.  Black hole is black circle
at center. Color scale is same as in figure~\ref{density}. }
\label{densityzoom}
\end{figure}

The Poynting-dominated jet forms as the differential rotation of the disk
and the frame-dragging of the black hole induce a significant toroidal
field that launches material away from the black hole by the same
force described in \citet{mckinney2005b}.

A coronal outflow is also generated between the disk and
Poynting-dominated jet.  In this model the coronal outflow has
$\Gamma_\infty\sim 1.5$.  The coronal-funnel boundary contains shocks
with a sonic Mach number of $M_s\sim 100$.  The inner-radial interface
between the disk and corona is a site of vigorous reconnection due to
the magnetic buoyancy and convective instabilities present there.
These two parts of the corona are about $100$ times hotter than the
bulk of the disk.  Thus these coronal components are a likely sites
for Comptonization and nonthermal particle acceleration.

Figure~\ref{density} and figure~\ref{densityzoom} show the final time
($t=1.4\times 10^4t_g$) log of rest-mass density and magnetic field
projected on the Cartesian z vs. x plane.  For the purposes of
properly visualizing the accretion flow and jet, we follow
\citet{mw99} and show both the negative and positive $x$-region by
duplicating the axisymmetric result across the vertical axis.  Color
represents $\log(\rho_0/\rho_{0,disk})$ with dark red highest and dark
blue lowest.  The final state has a density maximum of $\rho_0\approx
2\rho_{0,disk}$ and a minimum of $\rho_0\sim 10^{-13}\rho_{0,disk}$ at
large radii.  Grid zones are not smoothed to show grid structure.
Outer radial zones are large, but outer $\theta$ zones are below the
resolution of the figure.

Clearly the jet has pummelled its way through the surrounding medium,
which corresponds to the stellar envelope in the collapsar model. By
the end of the simulation, the field has been self-consistently
launched in to the funnel region and has a regular geometry there. In
the disk and at the surface of the disk the field is curved on the
scale of the disk scale height.  Within $r\lesssim 10^2r_g$ the funnel
field is ordered and stable due to the poloidal field dominance.
However, beyond $r\sim 10-10^2r_g$ the poloidal field is relatively
weak compared to the toroidal field and the field lines bend and
oscillate erratically due to pinch instabilities.  The radial scale of
the oscillations is $10^2 r_g$ (but up to $10^3r_g$ and as small as
$10r_g$), where $r\sim 10r_g$ is the radius where poloidal and
toroidal field strengths are equal.  By the end of the simulation, the
jet has only fully evolved to a state independent of the initial
conditions at $r\approx 5\times 10^3 r_g$, beyond which the jet
features are a result of the tail-end of the initial launch of the
field.  The head of the jet has passed beyond the outer boundary of
$r=10^4 r_g$.  Notice that the magnetic field near the black hole is
in an X-configuration.  This is due to the BZ-effect having power
$P_{jet}\propto \sin^2{\theta}$, which vanishes at the polar axis.
The X-configuration is also related to the fact that the disk+corona
is collimating the Poynting-dominated jet.  The field is mostly
monopolar near the black hole, and such field geometries {\it
decollimate} for rapidly rotating black holes in force-free
electrodynamics \citep{kras05}.

\begin{figure}
\includegraphics[width=3.3in,clip]{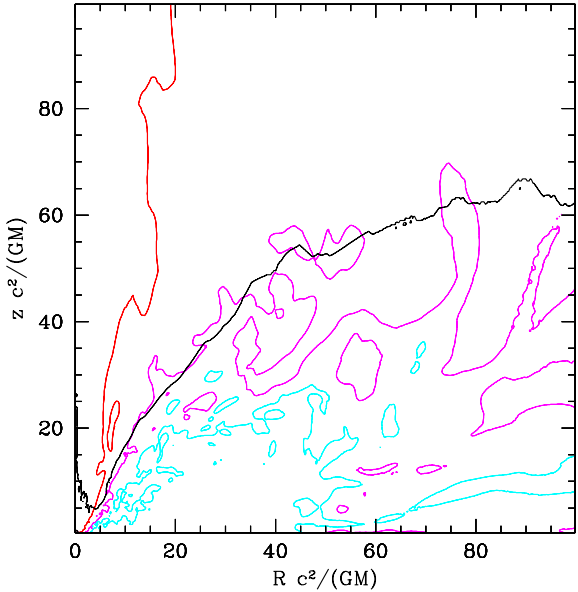}

\caption{ Contours for $t\approx 1500t_g$ and an outer scale of
$r\approx 10^2r_g$, where the disk-corona boundary is a cyan contour
where $\beta\approx u/b^2=3$, the corona-wind boundary is a magenta
contour where $\beta=1$, and the jet-wind boundary is a red contour
where $b^2/(\rho_0 c^2)=1$.  Black contour denotes boundary beyond
which material is unbound and outbound (wind + jet).  Beyond $r\approx
10r_g$ the jet undergoes poloidal oscillations due to a toroidal pinch
instability. }
\label{linesi}
\end{figure}

\begin{figure}
\includegraphics[width=3.3in,clip]{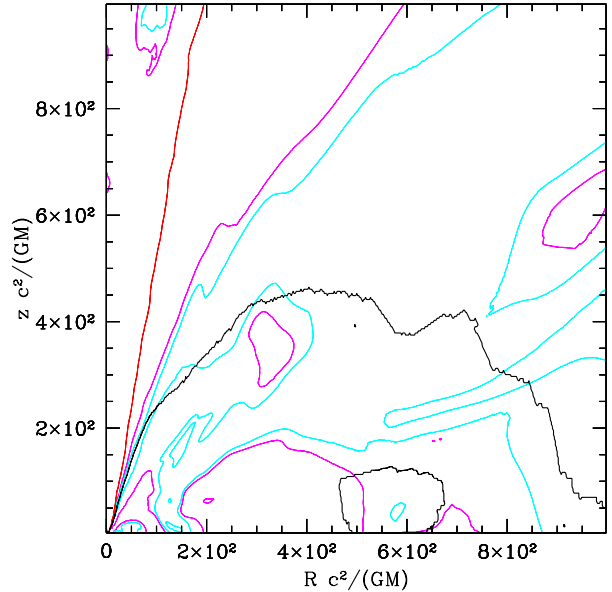}

\caption{ Same as figure~\ref{linesi} but for $t\approx 1.4\times
  10^4t_g$ and an outer scale of $r\approx 10^3r_g$.  Disk wind leads
  to broad coronal outflow.  Jet remains collimated to a fixed opening
  angle. By $r\gtrsim 100r_g$, pinch instabilities subside once
  magnetic energy converted into thermal energy and supports
  jet. Residual slow dense and fast diffuse patches from the
  instability are present in the jet, such as the slower and cooler
  dense blob shown at top left corner of figure. }
\label{lineso}
\end{figure}

\begin{figure}
\includegraphics[width=3.3in,clip]{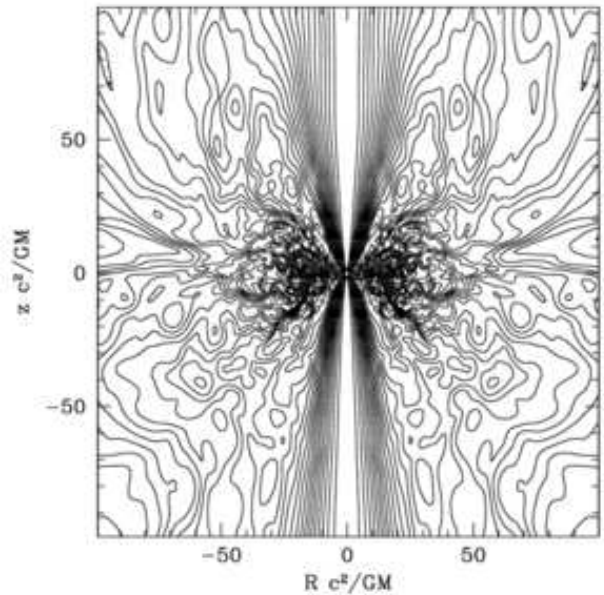}

\caption{Field geometry near black hole for $t\approx 1500t_g$ during
  phase of strong disk turbulence. }
\label{iaphizoom}
\end{figure}

Figures~\ref{linesi} and~\ref{lineso} show the energy structure of the
disk, corona, and jet.  This figure is comparable to the left panel of
figure 2 in \citet{mg04}.  Figure~\ref{linesi} shows contours for
$t\approx 1500t_g$ and an outer scale of $r=10^2r_g$, while
figure~\ref{lineso} shows contours for $t\approx 1.4\times 10^4t_g$
and an outer scale of $r=10^3r_g$. The disk-coronal boundary is
represented as a cyan contour where $\beta\equiv p_g/p_b \equiv
2(\gamma-1)u/b^2 = 3$, the coronal-wind boundary as a magenta contour
where $\beta=1$, and the jet-wind boundary as a red contour where
$b^2/(\rho_0 c^2)=1$.  The black contour denotes the boundary beyond
where material is unbound and outbound (wind or jet).  Beyond
$r\approx 10r_g$ the jet undergoes poloidal oscillations due to
toroidal pinch instabilities, which subside by $r\approx 10^2-10^3r_g$
where the jet is thermally supported.  At large scales, the cyan and
magenta contours closer to the equatorial plane are not expected to
cleanly distinguish any particular structure.

Figure~\ref{iaphizoom} shows the disk, corona, and jet magnetic field
structure during the turbulent phase of accretion at $t\approx
1500t_g$.  Compared to figure~\ref{densityzoom}, this shows the
turbulence in the disk, but is otherwise similar.  The jet, disk, and
coronal structures remains mostly unchanged at late times despite the
decay of disk turbulence.  That is, the current sheets in the disk do
not decay and continue to support the field around the black hole that
leads to the Blandford-Znajek effect.  The corona thickness and radial
extent do not require the disk turbulence, which only adds to the
time-dependence of the coronal outflow.

\begin{figure}
\subfigure{\includegraphics[width=3.3in,clip]{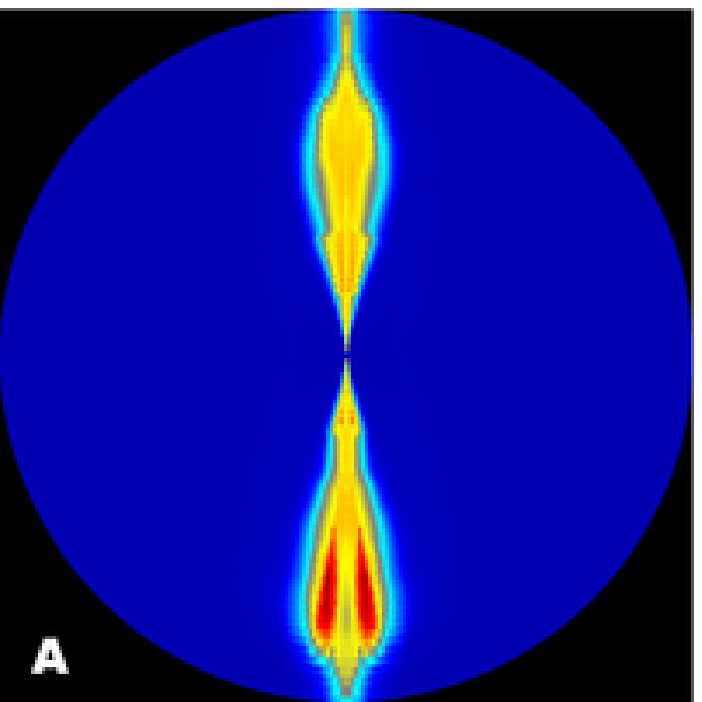}}
\subfigure{\includegraphics[width=3.3in,clip]{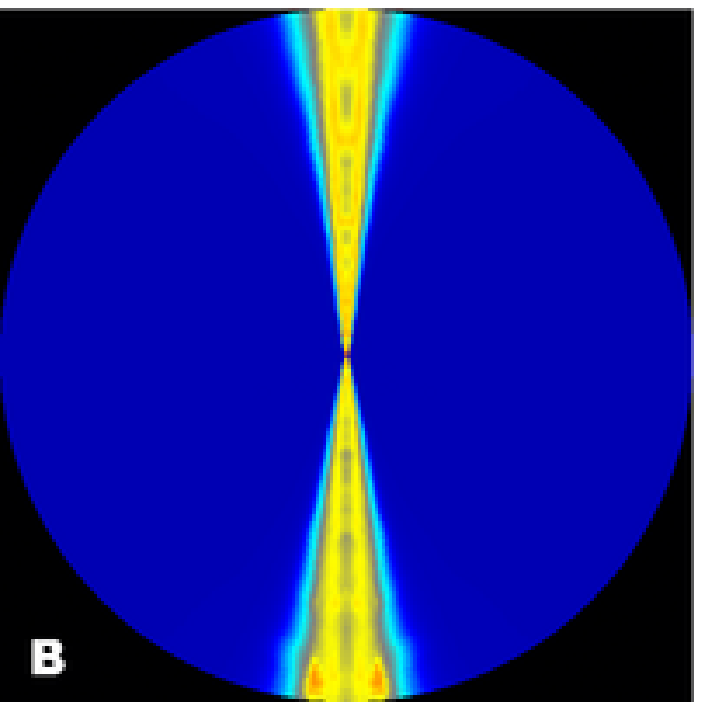}}

\caption{Jet becomes conical at large radii with a core half-opening
angle $\theta_j\approx 5^\circ$. Plot of $\log{\Gamma_\infty}$ with
red highest ($\Gamma_\infty\sim 10^4$) and blue lowest.  Yellow is
$\Gamma_\infty\sim 10^2 - 10^3$. Panel (A) has outer scale of $r=10^4
GM/c^2$.  Panel (B) outer scale $r=10^3 GM/c^2$.  Jet is independent
of initial conditions by $r\approx 5\times 10^3r_g$.}
\label{gammainf}
\end{figure}

\begin{figure}

\includegraphics[width=3.3in,clip]{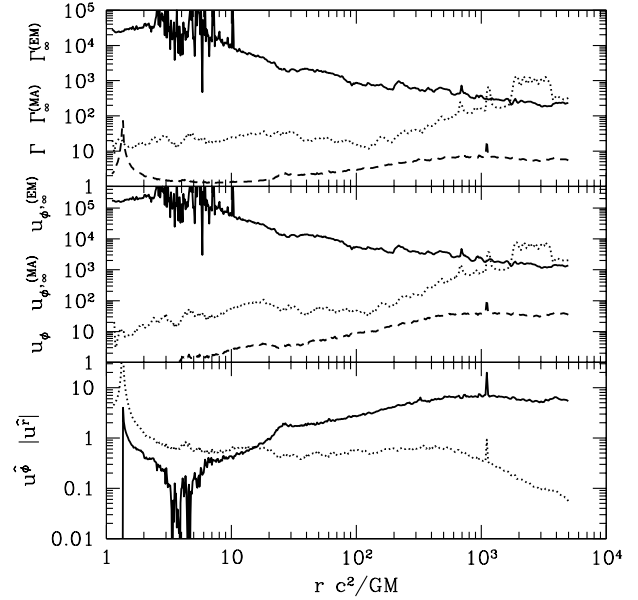}

\caption{A radial cross section along a mid-level field of the jet
showing the velocity structure. Top panel shows formation of magnetic
fireball where matter (MA) and electromagnetic (EM) energy flux per
unit rest-mass flux has $\Gamma^{(MA)}_\infty\sim
\Gamma^{(EM)}_\infty$.}
  \label{gammas}
\end{figure}

Figure~\ref{gammainf} shows a color plot of $\Gamma_\infty$, where red
is highest and reaches up to $\Gamma_\infty \sim 10^3 - 10^4$ and
yellow has $\Gamma_\infty\sim 10^2 - 10^3$.  Panel (A) shows outer
scale of $r=10^4 r_g$, while panel (B) shows outer scale of
$r=10^3r_g$ with same color scale.  The inner-radial region is not
shown since $\Gamma_\infty$ is divergent near injection region where
ideal-MHD breaks down.  Different realizations (random seed of
perturbations in disk) lead to up to about $\Gamma_\infty\sim 10^4$ as
shown for the lower pole in the color figures.  This particular model
was chosen for presentation for its diversity between the two polar
axes.  The upper polar axis is fairly well-structured, while the lower
polar axis has undergone an atypically strong magnetic pinch
instability.  Various realizations show that the upper polar axis
behavior is typical, so this is studied in detail below.  The strong
hollow-cone structure of the lower jet is due to the strongest field
being located at the interface between the jet and envelope, and this
is related to the fact that the BZ-flux is $\propto \sin^2{\theta}$,
which vanishes identically along the polar axis.  It is only the
disk+corona that has truncated the energy extracted, otherwise the
peak power would be at the equator.

Figure~\ref{gammas} shows the velocity structure of the
Poynting-dominated jet along a mid-level field line.  The top panel
shows the late-time time-averaged values of $\Gamma^{(EM)}_\infty$
(solid line), $\Gamma^{(MA)}_\infty$ (dotted line), and
$\Gamma=u^{\hat{t}}[{\rm BL-coords}]$ (dashed line) as a function of
radius along a mid-level field line.  For $2\lesssim r\lesssim 10$,
the value of $\Gamma^{(EM)}_\infty$ is highly oscillatory.  This shows
the location of the stagnation surface, where the poloidal velocity
$u^p=0$, is temporally variable.  The stagnation surface varies
between $2\lesssim r\lesssim 10$, within the range studied in prior
sections. This is where ideal-MHD approximation is breaking and thus
at any moment the value of $\Gamma^{(EM)}_\infty$ nearly diverges.
Notice that while at $r\lesssim 10^3r_g$ the Poynting flux dominates,
the Poynting flux energy is slowly converted into enthalpy flux due to
nonlinear time-dependent shock heating.  Shocks are expected beyond
the magneto-fast surface (see, e.g., \citealt{bt05}) and here are due
to pinch instabilities \citep{e93,begelman1998}.  For $r\gtrsim
10^3r_g$, the enthalpy flux and Poynting flux are in equipartition.

Thus a ``magnetic fireball'' has developed and the terminal Lorentz
factor is of order $\Gamma_\infty\sim 400$ for this choice of flow
line.  The Lorentz factor by $r\sim 10^4$ is still only $5\lesssim
\Gamma\lesssim 10$, with smaller patches with $\Gamma\sim 25$.  This
implies there will be an extended acceleration region where the
magnetic fireball loses energy during adiabatic expansion. These
results should also help motivate the boundary conditions used in jet
simulations \citep{aloy00,zwm03,zwh04}, which sometimes start out with
$\Gamma\sim 50$ and $u/\rho_0\sim 150$ near the black hole.  Such a
high enthalpy only occurs by $10^3r_g$ in our simulations.  Their
simulations are somewhat indicative of the Lorentz factor growth
beyond our simulated time and radial range.  This suggests that
$\Gamma\sim 100$ should occur by $r\sim 10^6r_g$ and that $\Gamma\sim
10^3$ should occur by $r\sim 10^7r_g$.  This is sufficient to avoid
the compactness problem.  A simulation of the acceleration region is
left for future work.

The next lower panel of figure~\ref{gammas} shows the electromagnetic
contribution to the specific angular momentum at infinity ($L=\Phi
B_\phi$) (solid line), the matter contribution to the specific angular
momentum at infinity ($h u_\phi$) (dotted line), and the specific
angular momentum $u_\phi$ (dashed line).

The bottom panel of figure~\ref{gammas} shows the orthonormal radial
($u^{\hat{r}}=\sqrt{g_{rr}} u^r$) (solid line) and $\phi$
($u^{\hat{\phi}}=\sqrt{g_{\phi\phi}} u^\phi$) (dotted line)
4-velocities in Boyer-Lindquist coordinates.  The directed motion
becomes relativistic, while the $\phi$-component of the 4-velocity
becomes sub-relativistic.  However, the rotation is still large enough
that it may stabilize the jet against $m=1$ kink instabilities, as
shown in the nonrelativistic case \citep{nakamura2004}.

\begin{figure}
\includegraphics[width=3.3in,clip]{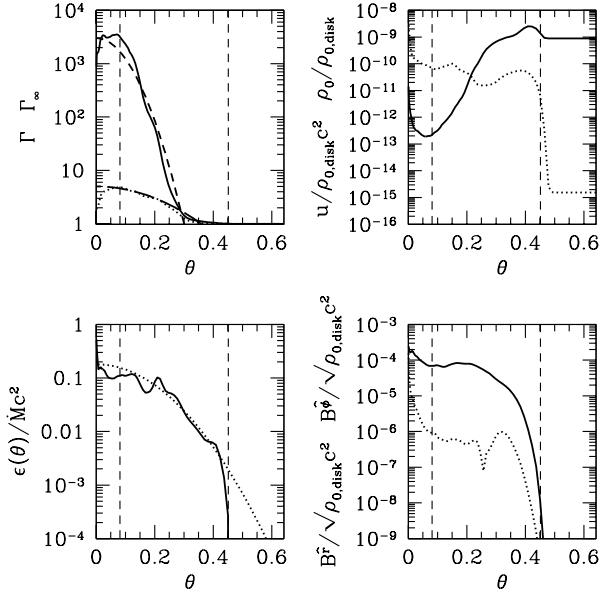}

\caption{ A $\theta$ cross-section of the jet at $r=5\times
10^3GM/c^2$ showing the velocity, density, energy, and magnetic
structure in Gauss. Velocity and energy structure show Gaussian fits
as dashed lines.  Vertical dashed lines marks outer and core jet. }
  \label{jetstruct}
\end{figure}

\begin{figure}

\includegraphics[width=3.3in,clip]{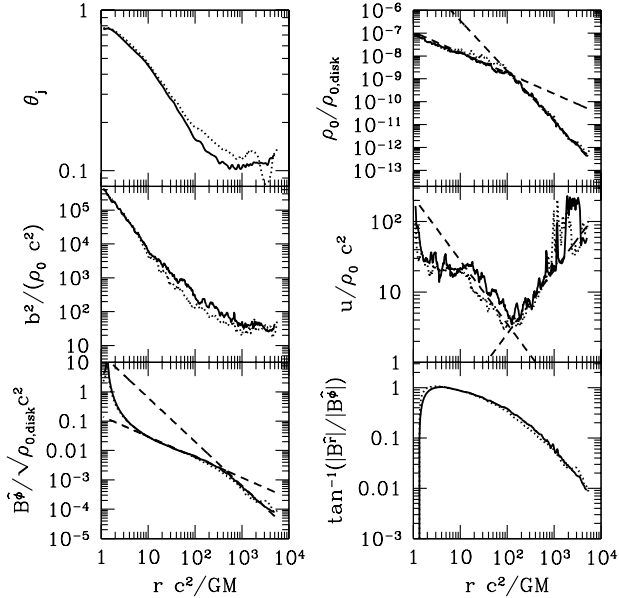}

\caption{ A radial cross-section of both poles of (solid and dotted
lines) jet along a mid-level field line in the jet showing the jet
collimation, density, and magnetic structure ($B^{\hat{\phi,r}}$ in
Gauss). Overlapping dashed lines are fits. }
  \label{jetstruct2}
\end{figure}

Figure~\ref{jetstruct} shows the $\theta$ cross-section of the jet at
$r=5\times 10^3 r_g$ for the upper polar axis according to
figures~\ref{density}, ~\ref{densityzoom}, and~\ref{gammainf}.  This
is a location by which the jet has stabilized in time, where farther
regions are still dependent on the initial conditions.  The hot fast
``core'' of the jet includes $\theta\lesssim 5^\circ$ at this radius
and is marked by the left dashed vertical line.  Also, it is useful to
notice that $\theta\approx 0.14$ corresponds to the radius obtained
for the ``mid-level'' field line shown for radial cross-sections in
figures~\ref{gammas} and~\ref{jetstruct2}.

The region within $\theta\lesssim 27^\circ$ is an expanded cold slow
portion of the jet.  It is possible that the gamma-ray burst photons
are due to Compton drag from soft photons emitted by this jet
``sheath'' dumping into the faster spine
\citep{bs87,ghis00,lazzati2004}.  Based upon the data fits described
below for $r\gtrsim 120r_g$ and a radiation-dominated plasma, the
outer sheath's ($\theta\approx 0.2$) seed photon temperature as a
function of radius is
\begin{equation}
T_{\gamma,seed}\sim 50{\rm keV} \left(\frac{r}{5\times 10^3r_g}\right)^{-1/3}
\end{equation}
These seed photons can be upscattered by the jet and produce a GRB and
high energy components.  A self-consistent study of a Compton dragged
jet that determines $\Gamma$ and the emission energy is left for
future work.

The right dashed vertical line marks the boundary between the last
field line that connects to the black hole.  Beyond this region is the
surrounding infalling medium.  The upper-left panel shows
$\Gamma_\infty$ (solid line), $\Gamma$ (dotted line) and Gaussian fits
to these two as dashed and long-dashed overlapping lines.  The lower
left panel shows the angular energy structure of the jet where
$\epsilon(\theta)\equiv -r^2 T^r_t\approx r^2 \rho_0 u^r
\Gamma_\infty$ with an overlapping Gaussian fit as a dotted line.  The
right panels show the density and magnetic structure of the jet.  The
upper right panel shows the density (solid line) and internal energy
density (dotted line).  The lower right panel shows the orthonormal
Boyer-Lindquist coordinate toroidal (solid line) and radial (dotted
line) field components.

This jet structure is weakly due to an interaction with the
surrounding medium, and primarily due to internal evolution of the
jet.  Clearly the slower jet envelope is non-negligible, giving
credence to the universal structured jet (USJ) model, but see
\citet{lamb2004} and see \citet{zdlm04,lr04}.  Gaussian jets have been
used to explain a universal connection between GRBs and x-ray flashes
\citep{zdlm04}. In most of our simulated models the jet structure is
Gaussian with $\epsilon(\theta)= \epsilon_0
e^{-\theta^2/2\theta_0^2}$, where $\epsilon_0\approx 0.18$ and
$\theta_0\approx 8^\circ$, within the range that they suggest fits
observations.  The total luminosity per pole is $L_j\approx
0.023\dot{M}_0c^2$, where $10\%$ of that is in the ``core'' peak
Lorentz factor region of the jet within a half-opening angle of
$5^\circ$.  Also, $\Gamma_\infty$ is approximately Gaussian with
$\Gamma_{\infty,0}\approx 3\times 10^3$ and $\theta_0\approx
4.3^\circ$.  Also, $\Gamma$ is approximately Gaussian with
$\Gamma_0\approx 5$ and $\theta_0\approx 11^\circ$.  This can be
folded into various models, such as the probability of observing
polarised emission in Compton drag emission models
\citep{ghis00,lazzati2004}.

For the collapsar model the above gives $L_j\approx 4\times
10^{51}\ergps$ for this $j=0.9375$ black hole model.  This is
approximately equal to the luminosity emitted by the black hole plus
the energy in pairs produced by annihilating neutrinos.  Not all of
this energy can be observed in $\gamma$-rays to obtain $E_j\approx
2\times 10^{51}\erg$ for a $30$ second event \citep{zdlm04}, unless
the black hole has spin $j\approx 0.6$ for most of the burst, which
would suggest little spin evolution that requires a much thicker disk
\citep{gsm04} and so less neutrino emission and annihilation
efficiency.  In \citet{mckinney2005b}, we found that $\Gamma_\infty$
is weakly dependent on the black hole spin for $a\gtrsim 0.6$, so the
Lorentz factor is likely unaffected by such changes.

It is more probable that the emission in $\gamma$-rays is inefficient
or only a fraction of the total energy is observed (i.e. this requires
$\lesssim 2\%$ for each if sole effect) ; or the true black hole mass
accretion rate is lower than predicted by current collapsar models
(i.e. $\dot{M}_0\sim 0.002\msun s^{-1}$ rather than $\dot{M}_0\sim
0.1\msun s^{-1}$ if sole effect).  For the standard collapsar model,
if one only observed the core of the jet, which has only $10\%$ of the
luminosity, then an efficiency of converting to $\gamma$-rays of
$\approx 20\%$ is required to fit the model of \citet{zdlm04}.

Notice that the disk thickness (and so mass accretion rate) may
strongly determine the collimation angle.  In higher mass accretion
rate systems, the disk is thicker than described here, and the final
opening angle may be smaller.  Also, if the mass accretion rate is
much lower, the disk is also thicker \citep{kohri2005}.  The collapsar
type model gives the thinnest disk near the black hole and small
changes in the mass accretion rate lead to a small change in the disk
thickness \citep{kohri2005}.  Also, for the relevant black hole spins,
the spin only weakly determines the final collimation angle
\citep{mg04}.  The universal jet model requires quite large opening
angles that would not be supported by invoking stellar (and so black
hole) spin dependence.  It would also not be supported by invoking
mass accretion rate dependence, since a system with a higher or lower
accretion rate actually has a thicker disk \citep{kohri2005} and so
stronger collimation.

There is a weak dependence on the density of the stellar envelope
compared to that seen in relativistic hydrodynamic simulations of jets
\citep{aloy00,zwm03,zwh04}, since here the confinement is mostly
magnetic.  During this energetic phase of the jet, the medium plays
little role in shaping the core of the jet, as has been suggested for
other reasons \citep{lazzati2005}.

The peak Lorentz factors are obtained in a narrow region within a core
with half-opening angle of $\theta\approx 5^\circ$.  Notice as well
that the very inner core of the jet is denser and slower, a feature
predicted in the theoretical discussion of \citet{mckinney2005b}.
This feature is simply a result that the Blandford-Znajek flux is
$\propto \sin^2{\theta}$.  Most of the Poynting energy is initially at
the interface between the jet and surrounding medium, until the jet
internally evolves to collimate into a narrower inner core.  Notice
that in \citet{mckinney2005b}, we predicted $\Gamma_\infty\sim 1000$
for $j=0.9$, where the simulated results show a peak
$\Gamma_\infty\approx 3\times 10^3$ and typical $10^2\lesssim
\Gamma_\infty \lesssim 10^3$. This is in reasonable agreement with
analytic estimates given the jet structure was not taken into account.

Figure~\ref{jetstruct2} shows the collimation angle along a mid-level
field line in the Poynting-dominated jet, along the same field line as in
figure~\ref{gammas}, and for both poles of the simulation.  Generally
the two poles are similar but not necessarily identical.  Many
simulations that were performed show at least one polar jet develop
some pinch instabilities.  In the model shown, one side develops very
strong instabilities by the end of the simulation.  The development of
the pinch instability mostly affects the far-radial Lorentz factor
($\Gamma$ and $\Gamma_\infty$ --- and so densities).  The value of
$\Gamma$ is up to 5 times larger in pinched regions than non-pinched
regions, while $\Gamma_\infty$ can be up to a factor 10 times larger
in pinched regions compared to non-pinched regions.

In figure~\ref{jetstruct2}, for $7r_g\lesssim r\lesssim 100r_g$ there
is a region of collimation slightly faster than logarithmic.  For the
field line closest to the coronal wind (not shown) starting at
$\theta_j\approx 1.0$, collimation is logarithmic with
\begin{equation}
\theta_j\approx 1 \left(\frac{r}{2.8r_g}\right)^{-1/3}.
\end{equation}
Closer to the polar axis the collimation is faster.  For the field
line starting at $\theta_j\approx 0.3$, approximately
\begin{equation}
\theta_j\propto r^{-2/5}
\end{equation}
up to $r\sim 10^2r_g$. The inner-radial collimation is due to
confinement by the disk+corona, while in the outer-radial range the
coronal outflow collimates the Poynting flow.  The corona is likely
required for the Poynting-dominated jet collimation, since without a
disk, a monopole-like field near the black hole actually decollimates
at higher black hole spin \citep{kras05}.  Far from the coronal
outflow, the internal jet is collimated by internal hoop stresses.
Note that the classical hoop-stress paradigm that jets can
self-collimate is not fully tested here, but these results suggest
that collimation is in large part due to, or at least requires, the
coronal outflow.

Notice from figure~\ref{gammas} that there is little acceleration
beyond $r\sim 10^2r_g$ while the magnetic fireball is forming.  It is
well known that magnetic acceleration requires field lines to
collimate \citep{begli94}.  Once the flow is pseudo-conical the flow
is unable to transfer magnetic energy into kinetic energy and proceeds
to convert it into enthalpy flux.  After $r\gtrsim 10^3 r_g$ the flow
oscillates around a pseudo-conical asymptote with a mean half-opening
angle of $\theta_j\sim 5^\circ$.  The opening angle has been found to
be weakly dependent on the black hole spin \citep{mg04} and, as found
in this paper, the details of the injection model.  This resulting
pseudo-conical asymptote results once the magnetic energy no longer
dominates and the coronal wind no longer helps collimate the flow.
This opening angle is in line with expectations built around afterglow
achromatic break measurements and estimates of the opening angle based
upon energy of the burst.  During the evolution of the burst, the jet
may continue to collimate due to a more extended coronal outflow.
This would be observed as an overall hard-to-soft evolution of
$\gamma$-rays since one is gradually exposed to the slower edges of
the jet since one likely does not observe exactly emission along the
core center.  The additional loading of neutrons from the coronal
outflow may also lead to a hard-to-soft evolution of the jet as the
overall Lorentz factor decreases over the duration of the burst.

Figure~\ref{jetstruct2} shows the toroidal field and pitch angle,
which shows that eventually the toroidal field completely dominates
the poloidal field, and the toroidal field remains ordered.  Thus,
large polarizations up to $60\%$ are possible \citep{gk03}.  The inner
radial toroidal field is well fit by
\begin{equation}
\frac{B^{\hat{\phi}}}{\sqrt{\rho_{0,disk}c^2}}[{\rm
Gauss}]=0.0023\left(\frac{r}{390r_g}\right)^{-0.7} .
\end{equation}
The outer radial field is well fit by
\begin{equation}
\frac{B^{\hat{\phi}}}{\sqrt{\rho_{0,disk}c^2}}[{\rm
Gauss}]=0.0023\left(\frac{r}{390r_g}\right)^{-1.5} .
\end{equation}
The transition radius is $r\approx 390r_g$. Notice that this is along
one mid-level field line, and the coefficients and power laws are
slightly different for each field line.  The pitch angle shows that
the field becomes toroidally dominated and at large radius the
magnetic loops have a pitch angle of $\lesssim 1^\circ$ by $10^3r_g$.
For regions not on the axis there is no possibility of reconnection
unless the flow is kink unstable \citep{dren02,ds02}.  Given the
regularity of the flow field and the lack of randomized field,
reconnection is {\it not} necessary or likely. Here, the pinch
instability drives a nonlinear coupling such as shocks that converts
the Poynting flux to enthalpy flux.  This mechanism should be
investigated in future work.

The toroidal field dominance leads to $m=0$ pinch magnetic
instabilities.  The $m=1$ kink mode may also operate, but studying the
3D jet structure is left for future work.  As such oscillations
develop, this leads to arbitrarily sized patches moving at arbitrary
relative velocities as in the internal shock model.  The typical size
of a patch is $\sim 100r_g$ to $\sim 10^3r_g$ in the lab frame along
the length of the jet.  For long-duration GRBs lasting about 30
seconds, the number of pulses should be no larger than the ratio of
the scale of the instability to the length of the event ($10^6 r_g$),
or about $100-1000$ pulses. Different realizations of the same
simulation show that the likelihood of an instability growing is
random, which could lead to the large variations in the number of
observed pulses and explain the diversity of observed pulses
\citep{np02}.  That is, some jets may have very few patches and so
pulses.  This type of instability is likely required to produce the
diversity of GRBs.  It is uncertain whether toroidal field instability
induced variability dominates pair creation loading variability due to
disk structure variability, which can be addressed by future work that
self-consistently treats the neutrino emission from the disk.

Figure~\ref{jetstruct2} also shows the rest-mass density per disk
density, internal energy density per rest-mass density, and the ratio
of (twice) the comoving magnetic energy density per unit rest-mass
density.  Notice that the density is close to that estimated in
\citet{mckinney2005b}, so those calculations are likely reasonable
approximations to the simulation results. The inner radial rest-mass
density is well fit by
\begin{equation}
\frac{\rho_0}{\rho_{0,disk}}=1.5\times
10^{-9}\left(\frac{r}{120r_g}\right)^{-0.9}.~~~{\rm (inner)}
\end{equation}
The outer radial rest-mass density is well fit by
\begin{equation}
\frac{\rho_0}{\rho_{0,disk}}=1.5\times
10^{-9}\left(\frac{r}{120r_g}\right)^{-2.2}.~~~{\rm (outer)}
\end{equation}
The transition radius is $r\approx 120r_g$.  Likewise, the inner
radial internal energy density is moderately fit by
\begin{equation}
\frac{u}{\rho_{0,disk}c^2}=4.5\times
10^{-9}\left(\frac{r}{120r_g}\right)^{-1.8}.~~~{\rm (inner)}
\end{equation}
The outer radial internal energy density is moderately fit by
\begin{equation}
\frac{u}{\rho_{0,disk}c^2}=4.5\times
10^{-9}\left(\frac{r}{120r_g}\right)^{-1.3}.~~~{\rm (outer)}
\end{equation}
The transition radius is $r\approx 120r_g$ and is due to the presence
of the thick disk at small radii.  Notice that this is along one
mid-level field line, and the coefficients and power laws are slightly
different for each field line.  One can compare this to the
``envelope'' density model used with
\begin{equation}
\frac{\rho_0}{\rho_{0,disk}}=8\times
10^{-8}\left(\frac{r}{120r_g}\right)^{-1.5}.~~~{\rm (envelope)}
\end{equation}
The envelope has little impact on the jet structure.

\begin{figure}
\includegraphics[width=3.3in,clip]{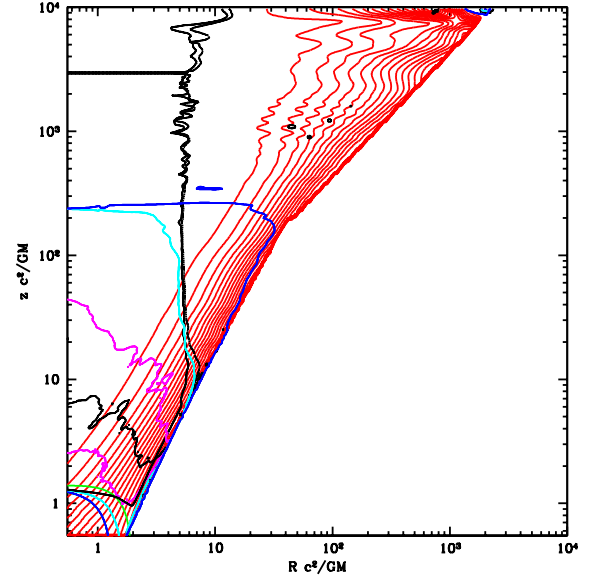}

\caption{Jet becomes superfast and toroidal field dominates at large
radii.  Stagnation surface time-dependent but stable.  Poynting-dominated
jet characteristic (and other interesting) surfaces. Red lines are
field lines. From $r=0$ outwards: Blue\#1: horizon + ingoing-fast ;
Cyan\#1: ingoing-\alf ; Black\#1: $B^{\hat{\phi}}=B^{\hat{r}}$ ;
Green\#1: ergosphere ; Purple\#1: ingoing-slow ; Black\#2: stagnation
surface where poloidal velocity $u^p=0$ ; Purple\#2: outgoing-slow ;
Cyan\#2: outgoing-\alf ; Black\#3: $B^{\hat{\phi}}=B^{\hat{r}}$ again
; Black\#4: light cylinder ; Blue\#2: outgoing-fast.}
  \label{wavespeeds}
\end{figure}

Figure~\ref{wavespeeds} shows the characteristic structure of the jet.
The figure shows a log-log plot of one hemisphere of the time-averaged
flow at late time.  In such a log-log plot, $45^\circ$ lines
correspond to lines of constant $\theta$.  Lines of constant spherical
polar $r$ are horizontal near the $z$-axis and vertical near the
$R$-axis.  The field lines are shown as red lines.  The disk and
coronal regions have been truncated with a power-law cutoff for
$r\lesssim 100r_g$ and a conical cutoff for larger radii.  The
inner-most blue line is the event horizon.  Clearly the field lines
are nearly logarithmic until $r\sim 100r_g$.  After this point the
field lines stretch out and oscillate around a conical asymptote.

Blue lines in figure~\ref{wavespeeds} show the ingoing and outgoing
fast surfaces.  Clearly the transition to a supercritical (superfast)
flow has occurred.  Within $r\lesssim 10r_g$ the plotting cutoff
creates the appearance that the fast surface and other lines terminate
along the cutoff.

The inner magenta line is the ingoing slow surface, while the outer
magenta line is the outgoing slow surface.  The inner cyan line is the
ingoing \alf~surface.  Energy can be extracted from the black hole if
and only if the \alf~surface lies inside the ergosphere \citep{tak90}.
The ergosphere is shown by the green line, which is indeed outside the
ingoing \alf~surface.  The outer cyan line is the outgoing
\alf~surface. The inner black line (next to the ergosphere) is the
surface where $B^{\hat{\phi}}=B^{\hat{r}}$ in Boyer-Lindquist
coordinates.  At larger radii there are two overlapping black lines.
One corresponds to, again, where the toroidal field is equal to the
poloidal field.  The other corresponds to the light ``cylinder''
surface.

Other small artifacts in the plot are a result of unsteady nature of
the flow.  However, the despite the unsteady nature of the flow the
characteristic structure is quite simple and relatively smooth in
appearance.  This indicates that the flow is mostly stationary.  The
stagnation surface is fairly unsteady, but stable.

All the results discussed in this paper are robust to increases or
decreases in numerical resolution.  Convergence testing has been
performed by choosing different models of the $\theta$ grid, including
different $Q_{jet}$ parameters that controls whether the disk or jet
is more resolved at small radii.  Another resolution of $256\times
256$ instead of $512\times 256$ was also chosen.  No qualitative
differences were found.  The interpolation scheme was also changed
from parabolic to linear and no qualitative differences for the
far-field jet region were found.

We find that the pair injection model weakly determines the bulk
structure of the flow at large radii.  Other injection models were
simulated by varying the total energy injected $\dot{e}_{inj}$,
fraction of rest-mass created $f_\rho$, fraction of internal energy
created $f_h$, fraction given to momentum energy $f_m$, and the
$\phi$-velocity of the injected particles.  Only $\dot{e}_{inj}$ and
$f_\rho$ strongly determine the bulk jet flow.

Most of these results are insensitive to the two different models of
the surrounding medium.  One model is a surrounding infall of material
and the other is an ``evacuated'' exterior region.  The jet structure
is negligibly broader or narrower in the evacuated case due to
magnetic confinement and collimation by the coronal outflow.

\subsection{Simulation as Applied to AGN and X-ray Binaries}

Notice that, quantitatively, all of these results can be simply
rescaled by density in the jet in order to apply to other GRB models,
AGN, and even approximately x-ray binary systems.  This is because by
the outer regions simulation, the jet has only reached about
$\Gamma\sim 5-10$ and $\Gamma_\infty\propto 1/\rho_{0,jet}$.

In particular, numerical models for M87 give the same results with
$2\lesssim \Gamma\lesssim 10$ patches by $r\sim 10^3r_g$.  The key
difference is that while the collapsar jet is very optically thick,
the M87 jet is not so most of the shock-heated energy is lost to
synchrotron radiation.  The same structured jet is formed, and is
useful to explain TeV BL Lac objects and radio galaxies \citep{ghis05}
and may help explain observations of blazars
\citep{bl95,ghis96,chiab00}.  The full opening angle of the jet core
of $\sim 10^\circ$ also agrees with the observations of the far-field
jet in M87 \citep{junor99}.  As discussed previously, the shocks are
synchrotron cooled and the heat generated does not contribute to
larger scale acceleration.  While self-consistent synchrotron cooling
should be included, these results suggest the shock zone (or ``blazar
zone'' for blazars) should be quite extended between $r\sim 10^2r_g$
and up to about $r\sim 10^4 r_g$.

Similarly, simulations were performed using an injection model for GRS
1915+105 giving patches with $1.5\lesssim \Gamma\lesssim 5$ by $r\sim
10^2 r_g$.  Due to the efficient pair-loading there is little extra
Poynting flux to convert to enthalpy flux or kinetic energy flux by
the time shocks occur at $r\sim 10^2 r_g$.  As discussed in
\citet{mckinney2005b}, the Poynting-lepton jet in the GRS 1915+105
model is possibly destroyed by Compton drag by disk photons and
synchrotron self-Compton and limited to at most $\Gamma\lesssim 2$,
but the jet could be completely destroyed by Compton drag.

\subsection{Summary of Fits}\label{summaryfit}

A summary of the fits along a fiducial field line is given.  Near
the black hole the half-opening angle of the full Poynting-dominated jet
is $\theta_j\sim 1.0$, while by $r\sim 120r_g$, $\theta_j\sim 0.1$.
This can be roughly fit by
\begin{equation}
\theta_j\sim \left(\frac{r}{r_g}\right)^{-0.4}~~~{\rm (inner)}
\end{equation}
for $r<120r_g$ and $\theta_j\sim 0.14$ beyond.  The core of the jet
follows a slightly stronger collimation with
\begin{equation}
\theta_j\propto r^{-2/5}
\end{equation}
up to $r<120r_g$ and $\theta_j\sim 0.09$ beyond.  Also, roughly for
M87 and the collapsar model, the core of the jet has
\begin{equation}
\Gamma_{bulk}\sim \left(\frac{r}{5r_g}\right)^{0.44}~~~{\rm (inner)}
\end{equation}
for $5<r\lesssim 10^3r_g$ and constant beyond for the M87 model if
including synchrotron radiation, while the collapsar model should
continue accelerating and the power law will truncate when most of the
internal and Poynting energy is lost to kinetic energy and the jet
becomes optically thin at about $r\sim 10^9r_g$ or internal shocks
take the energy away.  If the acceleration is purely thermal without
any magnetic effect, then $\Gamma\propto r$ \citep{mr97}.  However, it
is not clear how the equipartition magnetic field affects the
acceleration.  Roughly for GRS 1915+105 the core of the jet has
\begin{equation}
\Gamma_{bulk}\sim \left(\frac{r}{5r_g}\right)^{0.14}~~~{\rm (inner)}
\end{equation}
for $5<r\sim 10^3r_g$ and constant beyond, with no account for Compton
drag or pair annihilation.  Also, for any jet system the base of the
jet has $\rho_0\propto r^{-0.9}~~~{\rm (inner)}$ for $r\lesssim 120
r_g$ and $\rho_0\propto r^{-2.2}~~~{\rm (outer)}$ beyond.  For the
collapsar and M87 models
\begin{equation}
\frac{\rho_0}{\rho_{0,disk}}\sim 1.5\times 10^{-9}
\left(\frac{r}{120r_g}\right)^{-0.9}~~~{\rm (inner)}
\end{equation}
and
\begin{equation}
\frac{\rho_0}{\rho_{0,disk}}\sim 1.5\times 10^{-9}
\left(\frac{r}{120r_g}\right)^{-2.2}~~~{\rm (outer)} ,
\end{equation}
while for GRS 1915+105 the inner-radial coefficient is $10^{-5}$ and
outer is $6\times 10^{-3}$. For the collapsar model, the inner radial
internal energy density is moderately fit by
\begin{equation}
\frac{u}{\rho_{0,disk}c^2}=4.5\times 10^{-9}\left(\frac{r}{120r_g}\right)^{-1.8}~~~{\rm (inner)} .
\end{equation}
The outer radial internal energy density is moderately fit by
\begin{equation}
\frac{u}{\rho_{0,disk}c^2}=4.5\times 10^{-9}\left(\frac{r}{120r_g}\right)^{-1.3}~~~{\rm (outer)}.
\end{equation}
The transition radius is $r\approx 120r_g$.  For M87 the internal
energy is near the rest-mass density times $c^2$ until $r\sim 120r_g$
when the dependence is as for the collapsar case.  For GRS1915+105 the
internal energy is near the rest-mass density times $c^2$ until $r\sim
120r_g$ and then rises to about $2.5$ times the rest-mass density
times $c^2$.  The inner radial toroidal lab field is well fit by
\begin{equation}
\frac{B^{\hat{\phi}}}{\sqrt{\rho_{0,disk}c^2}}[{\rm
    Gauss}]=0.0023\left(\frac{r}{390r_g}\right)^{-0.7}~~~{\rm (inner)}
\end{equation}
for $5<r<390r_g$. The outer radial toroidal lab field is well fit by
\begin{equation}
\frac{B^{\hat{\phi}}}{\sqrt{\rho_{0,disk}c^2}}[{\rm
  Gauss}]=0.0023\left(\frac{r}{390r_g}\right)^{-1.5}~~~{\rm (outer)}
\end{equation}
for $r>390r_g$.

For the typical jet with no atypical pinch instabilities, the energy
and velocity structure of the jet follow
\begin{equation}
\epsilon(\theta)= \epsilon_0 e^{-\theta^2/2\theta_0^2} ,
\end{equation}
where $\epsilon_0\approx 0.18$ and $\theta_0\approx 8^\circ$.  The
total luminosity per pole is $L_j\approx 0.023\dot{M}_0c^2$, where
$10\%$ of that is in the ``core'' peak Lorentz factor region of the
jet within a half-opening angle of $5^\circ$.  Also, $\Gamma_\infty$
is approximately Gaussian
\begin{equation}
\Gamma_\infty(\theta)= \Gamma_{\infty,0} e^{-\theta^2/2\theta_0^2} ,
\end{equation}
where $\Gamma_{\infty,0}\approx 3\times 10^3$ and $\theta_0\approx
4.3^\circ$.  Also, $\Gamma$ is approximately Gaussian
\begin{equation}
\Gamma(\theta)= \Gamma_0 e^{-\theta^2/2\theta_0^2} ,
\end{equation}
where $\Gamma_0\approx 5$ and $\theta_0\approx 11^\circ$.  The outer
sheath's ($\theta\approx 0.2$) seed photon temperature as a function
of radius is
\begin{equation}
T_{\gamma,seed}\sim 50{\rm keV} \left(\frac{r}{5\times
  10^3r_g}\right)^{-1/3} .
\end{equation}

\section{Discussion}\label{discussion}

A collapsar-type GRMHD simulation with a neutrino annihilation model
and Fick diffusion model has been studied.  A self-consistent
Poynting-dominated jet is produced and the Lorentz factor at large
distances is $\Gamma_\infty\sim 10^3$ in the core of the jet, but an
equally important structured component exists with $\Gamma_\infty\sim
10^2$.  The Lorentz factor of the jet is determined by the
electron-proton loading by Fick diffusion of neutrons.  Notice that
estimates of the baryonic mass-loading from the optical flash of
GRB990123 suggest a baryon loading that gives $\Gamma\sim 10^3$
\citep{sr03}, which is in basic agreement with the findings here. The
half opening angle of the core of the jet is $\theta_j\sim 5^\circ$,
while there exists a significant structured component with a half
opening angle of $\theta_j\lesssim 25^\circ$.  It is likely that this
jet component survives traversing the remaining portion of the stellar
envelope ($r\sim 10^5r_g$, another decade in radius).

The jet at large distances is unstable to, at least, an $m=0$ pinch
instability due to the dominance of the toroidal field to the poloidal
field (see, e.g. \citealt{begelman1998}).  Here it is found that the
energy of the jet is patchy.  Typical realizations of the same model
have a random number of patches with $100\lesssim
\Gamma_\infty\lesssim 10^3$.  This likely results in a pulsed sequence
of magnetic fireballs that move at large relative Lorentz factors, as
required by the internal shock model \citep{mes02,gcg03,piran2005}.
The acceleration region and internal shock region was not simulated
since the dynamical range required is another $\sim 6$ orders of
magnitude in radius.

Some models assume the Poynting energy to be converted into radiation
far from the collapsing star by internal dissipation (see, e.g.,
\citealt{thompson94,mr97,sdd01,dren02,ds02,sbcp03,lpb03}) such as
magnetic reconnection. However, in our model, significant internal
dissipation occurs already by $10^3r_g$.

A very efficient model that reduces the compactness problem invokes
Compton drag (bulk Comptonization) to generate the emission, where the
stellar envelope or presupernova region provides soft photons
\citep{ghis00,lazzati2004}.  Our numerical model shows a jet structure
with a slow cold ``sheath'' necessary for supplying the seed photons,
so this process might explain the GRB emission process.

As in \citet{mg04}, a mildly relativistic Poynting-baryon jet is
launched from the inner edge of the disk with a half-opening angle of
roughly $16^\circ$ to $45^\circ$, however long-time study of this jet
component depends on sustaining long-time disk turbulence, which
cannot be simulated in axisymmetry.  For GRBs, this hot baryon-loaded
jet might play a significant role in the subsequent supernova (by
producing, e.g., $^{56}{\rm Ni}$) and be an interesting site for the
r-process \citep{kohri2005}.  The mass ejection rate is found to be
$\dot{M}_{0,coronal}\approx 0.03\dot{M}_0$, which for a $30$ second
event gives $0.1\msun$ of baryonic mass.  This may be sufficient to
power supernovae.  The Lorentz factor is about $\Gamma\sim 1.5-3$,
such as observed in a component of SN1998bw \citep{kulkarni1998}.

For GRBs, the picture that emerges is that Poynting flux is emitted
from the black hole at $r\sim r_g$ and is loaded with
electron-positron pairs within $r\lesssim 20r_g$.  A similar
mass-fraction of neutrons Fick-diffuses into the jet, and they
dominate the rest-mass once the temperature decreases to $T\lesssim
6\times 10^9$K by $r\sim 10r_g$.  Magnetic acceleration occurs over
$r\lesssim 10^2r_g$ up to $\Gamma\sim 10$.  Notice that this is in
contrast to \citet{mr97} who assumed the initial bulk Lorentz factor
is $\Gamma[r=r_H]\sim\Gamma_\infty^{(EM)}$, which assumed
instantaneous magnetic acceleration.

Once the toroidal field dominates the poloidal field, magnetic
instabilities develops around $r\sim 10-10^2 r_g$ which turns the jet
into an equipartition ``magnetic fireball'' by $r\sim 10^3 r_g$.  The
spatial structure of the magnetic fireball energy flux is patchy on
the instability scale of $r\sim 10^2 r_g$.  Simulations that
demonstrate how these features would later interact require much more
radial dynamic range.  Between $10^4 r_g \lesssim r \lesssim 10^8 r_g$
the flow accelerates due to adiabatic expansion with a radial
dependence up to $\Gamma\propto r$ \citep{mr97}, where the Poynting
flux provides a reservoir of magnetic energy that is continuously
shock-converted into thermal energy.

After passing the stellar surface at $r\sim 10^5 r_g$, the plasma
becomes transparent to $\gamma$-rays at $10^6 r_g\lesssim r \lesssim
10^7 r_g$, where possibly Compton drag of the sheath seed photons
produces the GRB emission \citep{ghis00,lazzati2004}.  By $r\sim
10^7$, patches of varying $\Gamma\sim 100-1000$ have reached their
terminal velocity.  Between $10^8 r_g \lesssim r \lesssim 10^{10} r_g$
the fireball is optically thin to Compton scattering, radiation
escapes, and pairs no longer annihilate but can continue to be
magnetically accelerated \citep{mr97}.  Within this same radial range,
internal shocks proceed to convert the remaining kinetic energy to
nonthermal $\gamma$-rays.  Beyond $10^{10}r_g$ a reverse shock may
occur and external shocks with the interstellar medium (ISM) occur.

Simulations of AGN were performed, which show similar $\Gamma$
vs. radius as the collapsar case.  That is, there is an early transfer
of Poynting flux to kinetic energy flux leading up to about
$\Gamma\sim 5-10$ by about $r\sim 10^3 r_g$ for an M87 model.  This is
consistent with the lack of observed Comptonization features in
blazars (see, e.g., \citealt{sikora2005}), although this is also
consistent with the fact that the optical depth to Comptonization is
low in such systems.

While prior work assuming $\Gamma_{bulk}\sim 10$ and $\theta_j\sim
15^\circ$ found that the BZ power is insufficient to account for
Blazar emission \citep{mt03}, as shown here the structure of the
Poynting-dominated jet is nontrivial.  The region with $\Gamma_{bulk}\sim
10$ is narrower with $\theta_j\sim 5^\circ$ and jet emission is likely
dominated by shock accelerated electrons with $\Gamma_e\gtrsim 100$
with an extended high energy tail.  This lowers the necessary energy
budget of the jet to be consistent with BZ power driving the jet.

\subsection{Theoretical Contemporaneous Comparisons}

Theoretical studies of ideal MHD jets focused on cold jets and the
conversion of Poynting energy directly to kinetic energy (see, e.g.,
\citealt{lcb92,begli94,dd02}).  Indeed, much of the work has focused
on self-similar models, of which the only self-consistent solution
found is suggested to be R-self-similar models \citep{vlah04}.
Unfortunately, such models result in only cylindrical asymptotic
solutions, which is apparently not what is observed in AGN jets nor
present in the simulations discussed in this paper.  The previous
section showed that some aspects of the jet are nearly self-similar
and that there is some classic ideal-MHD acceleration occurring in
this region.  However, once the toroidal field dominates the poloidal
field, the Poynting flux is converted into internal energy until they
reach equipartition.  The conversion is due to a time-dependent
nonlinear coupling, such as shocks, due to the magnetic pinch
instability.  Cold ideal-MHD jet models would have no way of
addressing this. This region also involves relatively rapid variations
in the flow, so stationary jet models would have difficulty modelling
this region.

Notice that GRMHD numerical models show that the magnetic field
corresponding to the Blandford-Znajek effect dominates all other
magnetic field geometries \citep{hirose04,mckinney2005a}.  For
example, there are no dynamically stable or important field lines
that tie the black hole to the outer accretion disk as in
\citet{uzdensky2005}.  All black hole field lines tie to large
distances or tie to the horizon-crossing accretion disk.  Also,
there are no field lines that connect the inner-radial accretion
disk and large distances.  The Blandford-Znajek associated magnetic
field is completely dominant, as shown in, e.g.,
figure~\ref{density} and figure~\ref{densityzoom}.

We have not directly included those effects of neutron diffusion that
induce a jet structure \citep{le2003}.  However, there is probably
strong mixing within the jet and the large-scale jet is probably not
affected by the details of the interface where neutron-diffusion
occurs.  Alternatively, the Gaussian structure we find may dominate
the power law structure they find since the Gaussian structure is due
to a strong internal electromagnetic evolution of the jet and matter
plays little role in setting the jet structure.

\subsection{Numerical Contemporaneous Comparisons}

\citet{mw99} evolve a presupernova core with a black hole inserted to
replace part of the core and evolve the nonrelativistic viscous
hydrodynamic equations of motion for various values of viscosity
coefficient ($\alpha$), mass accretion rate, nuclear burning, stellar
angular momentum, and some models have an injected energy at the poles
at some fixed energy rate.  For models in which they inject energy at
some specified rate, they find the baryon-contamination problem is
somewhat alleviated by forming a hot bubble. As shown in their figure
28, they find the jet has $u/(\rho_0 c^2)\sim 10$ and they suggest
that this corresponds to $\Gamma_\infty\sim 10$, insufficient to avoid
the compactness problem.  Their $\alpha=0.1$ model shows a significant
disk outflow, which over the duration of the GRB would yield $M\sim
\msun$ in mass that could power a supernova.  In their $\alpha=0.01$
model there are insignificant outflows.

Our MHD results are most comparable to their $\alpha=0.01$ model.  We
find a disk outflow yielding $M\sim 0.1\msun$, which may still be
sufficient to produce a supernova component.  We have
self-consistently evolved the jet formation and included an
approximate model for the pair creation physics.  In our case the
baryon-contamination problem is self-consistently avoided by magnetic
confinement of the jet against baryons, where only a small number of
neutrons Fick-diffuse into the jet and self-consistently determine the
Lorentz factor at large distances.  We find that neutrino annihilation
energy is probably dominated by energy from the black hole.  Notice
that we find $\Gamma_\infty\sim 100-1000$, which is much larger than
they find.  The difference between their and our model is the presence
of a magnetic field and a rotating black hole, which drive the
BZ-effect and a stronger evacuation of the polar jet region.

\citet{proga05} have performed nonrelativistic MHD simulations
of collapsars with a realistic equation of state.  They suggest MHD
accretion is able to launch, collimate, and sustain a strong Poynting
outflow, although they measure a jet velocity of $v\sim 0.2c$.  They
find the jet is Poynting flux dominated such that
$\Gamma_\infty\lesssim 10$, insufficient to avoid the compactness
problem.  The rotation of the black hole is crucial to generate a
sufficiently Poynting-dominated jet.  Similar fully relativistic
simulations by \citet{mizuno04b} have performed with a jet velocity of
only $v\sim 0.3c$, which is due to their choice of the ``floor'' model
in the polar regions and the short time of integration.

\citet{zwm03,zwh04} use a relativistic hydrodynamic model to simulate
jet propagation through the stellar envelope and subsequent breakout
through the stellar surface.  They assume a highly relativistic jet is
formed early near the black hole and they inject the matter with a
large enthalpy per baryon of about $u/\rho_0c^2\sim 150$.  They also
tune the injected energy to compare with observations \citep{frail01}.
In their view, variability is due to hydrodynamic instabilities
between the cocoon and core of the jet.  They find the stellar
envelope can collimate the flow.

We have self-consistently evolved the {\it formation} of the jet along
with the disk without having to assume a jet structure.  We suggest
that the injected energy per baryon is only $u/\rho_0c^2\lesssim 20$
unless super-efficient neutrino mechanisms are invoked.  We also find
that the self-consistent energy released is larger than observed,
suggesting an fairly inefficient generation of $\gamma$-rays and a
dominant lower $\Gamma$ jet component that may explain x-ray flashes.
We suggest magnetic instabilities due to pinch or kink modes dominate
the variability rather than hydrodynamic instabilities, which are
known to be quenched by magnetic confinement.  We find that the jet
structure is negligibly broader for models with no stellar envelope
due to magnetic confinement and we find that the outer portion of the
Poynting-dominated jet is collimated by the coronal outflow.

Many hydrodynamic jet propagation simulations have been studied (for a
review see \citealt{scheck2002}).  \citet{scheck2002} evolve
hydrodynamic relativistic hadronic and leptonic jets using a realistic
equation of state.  They find that the resulting morphology of the
jets are similar, despite the different composition.  This suggests
jet gross morphology is not a useful tool to differentiate jet
composition.  In our model, the most relativistic jet component is
lepton-dominated in AGN and x-ray binaries and baryon-dominated in GRB
systems.  Notice that magnetic confinement quenches most hydrodynamic
instabilities.

\citet{dv05b} evolve a fully relativistic, black hole mass-invariant
model, showing a jet with hot blobs moving with $\Gamma\sim 50$. A
primary result in agreement with the results here is that a patchy or
pulsed ``magnetic fireball'' is produced.  This is gratifying since it
suggests that the development of a ``magnetic fireball'' is not an
artifact of the numerical implementation but is a likely result of
shock heating.  We also agree in finding that the core of the jet is
hot and fast and is surrounded by a cold slow flow.  One difference is
that they say they seem to find the flow is cylindrically collimated
by $r\gtrsim 300r_g$, while we find nearly logarithmic collimation
until $r\sim 10^2 r_g$ and an oscillatory conical asymptote beyond.
Another difference is that they suggest temporal variability is due to
injection events near the black hole, while we suggest it is due to
pinch (or perhaps kink) instabilities at $r\gtrsim 10^2r_g$.

Another difference is that they find a larger value of $\Gamma$ at
smaller radius than we find.  Indeed, one should wonder why their
black hole mass-invariant model applies to only GRBs and not AGN or
x-ray binaries.  This is a result of their {\it much} lower ``floor''
density of $\rho_0\sim 10^{-12}\rho_{0,disk}$, which is not consistent
with self-consistent pair-loading or neutron diffusion loading. Also,
because they use a constant density floor, the jet region at large
radius must be additionally loaded with an arbitrary amount of
rest-mass.  The typical ``floor'' model adds rest-mass into the
comoving frame, but this artificially loads the jet with extra mass
moving at high $\Gamma$.  We suggest that rather than the acceleration
occurring within $r\lesssim 700r_g$ of their simulation, that
acceleration occurs much farther away before the emitting region at
$r\lesssim 10^9 r_g$.  In their $\Gamma\sim 50$ knots the gas has
$10^3\lesssim u/\rho_0c^2\lesssim 10^6$.  This implies that their
actual terminal Lorentz factor is on the order of $10^5\lesssim
\Gamma_\infty\lesssim 10^7$, which would imply that the external
shocks occur before internal shocks could occur.

\citet{kom05} study the BZ process and MHD Penrose process.  They
found that the prior ``jet'' results of \citep{koide2002}, who evolve
only for $t\sim 100t_g$, are only a transient phenomena associated
with the initial conditions.  They also suggest that there are serious
problems for the MHD-driven model of jets since they do not find jets
in their models except for contrived field geometries.  However, this
is potentially due to three effects.  First, their outer radius is
$r\sim 50r_g$, while here we study out to $r\sim 10^4r_g$.  The
magnetic acceleration only leads to a logarithmic increase in the
velocity, so a large radial range is required to observe relativistic
motion.  Second, magnetic acceleration requires collimating field
lines \citep{begli94}, and a disk or disk wind is probably required to
collimate the magnetic outflow \citep{okamoto1999,okamoto2000}.  The
``disk'' that forms in their models is relatively thin.  Third, to
avoid numerical errors, they limit their solution to $b^2/(\rho_0
c^2)\lesssim 100$.  Here we find that a self-consistent source of jet
matter from pair creation and neutron diffusion leads to $b^2/(\rho_0
c^2)\sim 10^5$ near the black hole.  These three effects are probably
why they find no relativistic jets.

\subsection{Limitations}

As has been pointed out by \citet{kom05}, nonconservative GRMHD
schemes often overestimate the amount of thermal energy produced.  All
GRMHD numerical models suffer from some numerical error.
Shock-conversion of magnetic energy to thermal energy is modelled by
HARM in the perfect magnetic fluid approximation with total energy
conserved exactly.  However, our numerical model may overestimate the
amount of magnetic dissipation in shocks, and so the shocks may more
slowly convert magnetic energy to thermal energy.  If this dissipation
was delayed until the $\gamma$-ray photosphere at $r\sim 10^7 - 10^8
r_g$, then those shocks could be directly responsible for the
$\gamma$-ray emission from GRBs.  Clarification of this issue is left
for future work.  However, we expect that toroidal field instabilities
drive efficient magnetic dissipation as shown in the numerical results
presented here.

In order to evolve for a longer time than simulated in this paper,
other physics must be included.  For long-term evolution of a GRB
model, one must include disk neutrino cooling, photodisintegration of
nuclei, and a realistic equation of state.  If one wishes to track
nuclear species evolution, a nuclear burning reactions network is
required.  For the neutrino optically thick region of the disk,
radiative transport should be included.  The self-gravity of the star
should be included to evolve the core-collapse.  This includes a
numerical relativity study of the collapse of a rotating magnetized
massive star into a black hole (see review by \citealt{st03}, \S 4.3).
The jet should be followed through the entire star and beyond
penetration of the stellar surface \citep{aloy00,zwm03,zwh04}.

As applied to all black hole accretion systems, some other limitations
of the numerical models presented include the assumption of
axisymmetry, ideal MHD, and a nonradiative gas.  The assumption of
axisymmetry is likely not important for the inner jet region since our
earlier results \citep{mg04} find quantitative agreement with 3D
results \citep{dhk03}.  The primary observed limitation of axisymmetry
appears to be the decay of turbulence \citep{cowling34}, which we
attempt to avoid by requiring a resolution that gives quasi-steady
turbulence for much of the simulation.  Also, the jet at large
distances has already formed by the time turbulence decays, and by
that time the jet at large radius is not in causal contact with the
disk.

When the toroidal field dominates the poloidal field, eventually $m=1$
kink instabilities and higher modes may appear (see, e.g.,
\citealt{nakamura2001,nakamura2004}).  Thus our models may
underestimate the amount of oscillation in the flow and the conversion
of Poynting flux to enthalpy flux.

The limitation of axisymmetry may also {\rm limit} the efficiency of
the Blandford-Znajek process.  The magnetic arrested disk (MAD) model
suggests that any accretion flow likely accumulates a large amount of
magnetic flux near the black hole.  This means the efficiency of
extracting energy would be higher
\citep{igumenshchev2003,narayan2003}.

We have neglected low-energy jet and disk radiative processes in the
numerical simulations.  Future work should include a model of
synchrotron radiation for an AGN model in order to simulate the
``jet'' emission to verify the claims made in \citet{mckinney2005b}
that the broad ``jet'' emission observed by \citet{junor99} is
actually the coronal outflow rather than the Poynting-dominated jet.

Also for AGN, a self-consistent simulation with synchrotron emission
would likely show the continuous loss of Poynting flux until the
synchrotron cooling timescale is longer than the jet propagation
timescale, which still suggests the jet Poynting flux is finally in
equipartition with the enthalpy flux.  This would suggest that the
shock-zone, and so emission region, is more extended ($r\sim 10^2 -
10^4r_g$) than the simulated shock-zone ($r\sim 10^2 - 10^3 r_g$).

For AGN and x-ray binaries, the radiatively inefficient disk
approximation, which assumes electrons couple weakly to ions, may not
hold.  If the electrons and ions eventually couple near the black
hole, then the disk might collapse into an unstable magnetically
dominated accretion disk (MDAF) \citep{meier2005}.  Like the MAD
model, this might drastically alter the results here, although it is
uncertain whether jets are actually produced under the conditions
specified by the MDAF model.

The single-fluid, ideal MHD approximation breaks down under various
conditions, such as during the quiescent output of x-ray black hole
binaries, where a two-temperature plasma likely forms near the black
hole as ions and electrons decouple (see, e.g.,
\citealt{ny95}). Resistivity plays a role in current sheets where
reconnection events may generate flares as on the sun, such as
possibly observed in Sgr~A \citep{genzel2003}.  Finally, radiative
effects may introduce dynamically important instabilities in the
accretion disk (e.g.  \citealt{gam98,bs03}).

\section{Conclusions}\label{conclusions}

Primarily two types of relativistic jets form in black hole (and
perhaps neutron star) systems.  The Poynting-dominated jet region is
composed of field lines that connect the rotating black hole to large
distances.  Since the ideal MHD approximation holds very well, the
only matter that can cross the field lines are neutral particles, such
as neutrinos, photons, and free neutrons.

In \citet{mckinney2005b}, we showed that the primary differences
between GRBs, AGN, and black hole x-ray binaries is the pair-loading
of the Poynting-dominated jet, a similar mass-loading by free neutrons
in GRB-type systems, the optical depth of the jet, and the synchrotron
cooling timescale of the jet.

In \citet{mckinney2005b}, we showed that for GRB-type systems the
neutron diffusion flux is sufficiently large to be dynamically
important, but small enough to allow $\Gamma\sim 100 - 1000$.  Beyond
$r\sim 10r_g$ many of the electron-positron pairs annihilate, so the
Poynting-dominated jet is dominated in mass by electron-proton pairs
from collision-induced neutron decay.  Most of the energy is provided
by the BZ effect instead of neutrino-annihilation.

In \citet{mckinney2005b}, we showed that for AGN and x-ray binaries,
the density of electron-positron pairs established near the black hole
primarily determines the Lorentz factor at large distances.
Radiatively inefficient AGN, such as M87, achieve $2\lesssim
\Gamma_\infty\lesssim 10$ and are synchrotron cooling limited.  The
lower the $\gamma$-ray radiative efficiency of the disk, the more
energy per particle is available in the shock-zone.  Radiatively
efficient systems such as GRS1915+105 likely have no Poynting-lepton
jet due to strong pair-loading and destruction by Comptonization by
the plentiful soft photons for x-ray binaries with optically thick
jets.  However, all these systems have a mildly relativistic,
baryon-loaded jet when in the hard-low state when the disk is
geometrically thick, which can explain jets in most x-ray binary
systems.

A GRMHD code, HARM, with pair creation physics was used to evolve many
black hole accretion disk models.  The basic theoretical predictions
made in \citet{mckinney2005b} that determine the Lorentz factor of the
jet were numerically confirmed.  However, Poynting flux is not
necessarily directly converted into kinetic energy, but rather
Poynting flux is first converted into enthalpy flux into a ``magnetic
fireball'' due to shock heating.  Thus, at large distances the
acceleration is primarily thermal, but most of that thermal energy is
provided by shock-conversion of magnetic energy.  In GRB systems this
magnetic fireball leads to thermal acceleration over an extended
radial range.  The jets in AGN and x-ray binaries release this energy
as synchrotron and inverse Compton emission and so the jet undergoes
negligible thermal acceleration beyond $r\sim 10^2-10^3r_g$.

Based upon prior theoretical \citep{mckinney2005b} and this numerical
work, basic conclusions for collapsars include:
\begin{enumerate}
  \item Black hole energy, not neutrino energy, typically powers GRBs.
  \item Poynting-dominated jets are mostly loaded by $e^- e^+$ pairs
  close to the black hole, and by $e^- p$ pairs for $r\gtrsim 10r_g$.
  \item BZ-power and neutron diffusion primarily determines Lorentz factor.
  \item Variability is due to toroidal field instabilities.
  \item Poynting flux is converted into enthalpy flux and leads to the
  formation of a ``magnetic fireball.''
  \item Patchy jet develops $10^2\lesssim \Gamma_\infty\lesssim 10^3$,
  as required by internal shock model.
  \item Random number of patches ($<1000$ for 30 second burst) and so
  random number of pulses.
  \item Energy structure of jet is Gaussian with $\theta_0\approx
  8^\circ$.
  \item Core of jet with $\theta_j\approx 5^\circ$ can explain GRBs.
  \item Extended slower jet component with $\theta_j\approx 25^\circ$
  can explain x-ray flashes.
  \item Coronal outflows with $\Gamma\sim 1.5$ may power supernovae
  (by producing, e.g., $^{56}{\rm Ni}$) with $M\sim 0.1\msun$
  processed by corona.
\end{enumerate}

Based upon prior theoretical \citep{mckinney2005b} and this numerical
work, basic conclusions for AGN or x-ray binaries include:
\begin{enumerate}
  \item Poynting-dominated jets $e^- e^+$ pair-loaded unless advect
  complicated field.
  \item $\gamma$-ray radiative efficiency, and so pair-loading,
  determines maximum possible Lorentz factor.
  \item Poynting-lepton jet is collimated with $\theta_j\approx 5^\circ$.
  \item Extended slow jet component with $\theta_j\lesssim 25^\circ$.
  \item For fixed accretion rate, variability is due to toroidal field
  instabilities.
  \item Poynting flux is shock-converted into enthalpy flux.
  \item In some AGN, shock heat in transonic transition lost to
  synchrotron emission and limits achievable Lorentz factor to
  $2\lesssim \Gamma\lesssim 10$  (e.g. in M87).
  \item Coronal outflows produce broad inner-radial jet features in AGN
  together with well-collimated jet component (e.g. in M87).
  \item In some x-ray binaries, Compton drag loads Poynting-lepton jets and limits
  Poynting-lepton jet to $\Gamma\lesssim 2$ or jet destroyed.
  \item In some x-ray binaries, Poynting-lepton jet optically thick and emits
  self-absorbed synchrotron.
  \item Coronal outflows have collimated edge with $\Gamma\lesssim 1.5$.
  \item Coronal outflows may explain all mildly relativistic and
  nonrelativistic jets in radiatively efficient systems (most x-ray
  binaries).
\end{enumerate}
For AGN and X-ray binaries, the coronal outflow collimation angle is
strongly determined by the disk thickness.  The above conclusions
regarding the collimation angle assumed $H/R\sim 0.2$ near the black
hole and $H/R\sim 0.6$ far from the black hole, while $H/R\sim 0.9$
(ADAF-like) is perhaps more appropriate for some systems.  The
sensitivity of these results to $H/R$ is left for future work.

\section*{Acknowledgments}

I thank Avery Broderick for an uncountable number of inspiring
conversations.  I also thank Charles Gammie, Brian Punsly, Amir
Levinson, and Ramesh Narayan, with whom each I have had inspiring
conversations.  I thank Scott Noble for providing his highly efficient
and accurate primitive variable solver.  I thank Xiaoyue Guan for
providing her implementation of parabolic interpolation.  This
research was supported by NASA-ATP grant NAG-10780 and an ITC
fellowship.

\begin{appendix}

\section{GRMHD Equations of Motion with Pair Creation}\label{GRMHD}

A single-component GRMHD approximation that accounts for baryon
conservation is summarized.  Leptons are assumed to be conserved in
the equation of state (EOS) and by accounting for free-streaming
neutrinos.  We assume an ideal gas EOS of relativistic particles and
neglect radiative transport since we assume all particles either
stream freely or are trapped in the fluid.  Alternatively, we assume
the initial conditions well-model the steady-state radiative
equilibrium.  We model the streaming photon/neutrino-annihilation into
pairs by injecting rest-mass and energy-momentum at an appropriate
fraction of the baryon density.

The black hole has a Kerr metric written in Kerr-Schild coordinates,
where the Kerr metric in Kerr-Schild coordinates and the Jacobian
transformation to Boyer-Lindquist coordinates is given in
\citet{mg04}.  We use Kerr-Schild rather than Boyer-Lindquist because
in Kerr-Schild coordinates the inner-radial boundary can be placed
inside the horizon and so out of causal contact with the flow.  It is
difficult to avoid interactions between the numerical inner boundary
and the jet when using Boyer-Lindquist coordinates.  This interaction
leads to excessive variability in the jet since the ingoing superfast
transition is not on the grid and then the details of the boundary
condition significantly impact the jet.  Numerical models of viscous
flows have historically had similar issues (see, e.g.,
\citealt{mg02}).

A single-component MHD approximation is assumed such that baryon
number is conserved up to a source term due to pair creation due to
either radiative annihilation or neutron diffusion, such that
\begin{equation}
(\rho_0 u^\mu)_{;\mu} = S_\rho ,
\end{equation}
where $\rho_0 \equiv m_b n_b$, $m_b\approx m_n$ the neutron mass,
$n_b=n_n+n_p$, and $m_p$ is the proton mass.  One may choose an
arbitrary mass weight to define $\rho_0$ in the single component fluid
approximation.

For a magnetized plasma the conservation of energy-momentum equations
are
\begin{equation}\label{EOM}
{T^{\mu\nu}}_{;\nu} = \left(T^{\mu\nu}_{\rm MA} + T^{\mu\nu}_{\rm
EM}\right)_{;\nu} = S_T^{\mu}.
\end{equation}
where $T^{\mu\nu}$ is the stress-energy tensor, which can be split
into a matter (MA) and electromagnetic (EM) part.  In the fluid
approximation
\begin{equation}
T^{\mu\nu}_{\rm MA} = (\rho_0 + u_g) u^\mu u^\nu + p_g P^{\mu\nu},
\end{equation}
with a relativistic ideal gas pressure $p_g=(\gamma-1) u_g$,
$\gamma=4/3$, and $P^{\mu\nu} = g^{\mu\nu} + u^\mu u^\nu$ is the
projection tensor.  Either $\gamma=5/3$ or $\gamma=4/3$ for either the
baryon-dominated component or lepton-dominated component does not
change the results described in section~\ref{results}.

\subsection{Injection Physics}

A detailed pair creation model would self-consistently determine the
distribution of rest-mass and energy-momentum injected, which is left
for future work.  In this paper, the rough results of
\citet{pwf99,mw99} are used to approximate the neutrino annihilation
and pair creation.  They determine the energy density creation rate in
pairs as measured by an observer at infinity ($\dot{e}_{inj}$).  A
monoenergetic, monomomentum injection of rest-mass and energy-momentum
is assumed.  That is, for a Lorentz invariant particle distribution
function of
\begin{equation}
f=\frac{dN}{dx^{\hat{1}}dx^{\hat{2}}dx^{\hat{3}}du^{\hat{1}}du^{\hat{2}}du^{\hat{3}}} ,
\end{equation}
the fluid equations are derived from the Boltzmann equation
\begin{equation}
\frac{df}{d\tau}=\frac{dn_{inj}}{d\tau}\delta[u^\mu - u^\mu_{inj}]
\end{equation}
for a particle creation density rate $dn_{inj}/d\tau$ in the comoving
frame of the existing fluid.  There is no collisional term in the
ideal case.  After mass-weighting and taking 4-velocity moments of the
Boltzmann equation one has $S_\rho=(\rho_{0,inj} u_{inj}^\mu)_{;\mu}$,
where $\rho_{0,inj}$ is the injected rest-mass density and
$u_{inj}^\mu$ is the 4-velocity of the injected particles.  Also,
$S_T^{\mu}={T^{\mu\nu}_{inj}}_{;\nu}$.  Here,
\begin{equation}
\frac{dn_{inj}}{d\tau}\equiv \frac{m_e}{m_b}\frac{dn_{e^-
    e^+,inj}}{d\tau}=\frac{1}{m_b}\frac{d\rho_{0,e^- e^+,inj}}{d\tau} ,
\end{equation}
so the fluid is still treated as a single component with a single mass
weight $m_b$.  As discussed in \citet{mckinney2005b}, this is a good
approximation since the late-time Poynting-dominated jet region is
dominated by lepton mass while the remaining region is dominated by
baryon mass.  The simultaneous advected flux of injected energy is
neglected, so then ${S_T}_t = \detg\dot{e}_{inj}$ is valid in
Boyer-Lindquist, Kerr-Schild, or modified Kerr-Schild coordinates.
This two-step approach (similar to operator splitting), where
particles are injected and {\it then} advected with the normal fluid,
is a good approximation for typical numerical integrations that use a
multi-step timestep approach. This is also a good approximation
because the injection region has a small 4-velocity beyond the
stagnation surface.  Also, the error in the injection approximations
being made is larger than the error in neglecting the advection term.
Thus
\begin{equation}\label{SRHO}
S_\rho = \partial/\partial t(\detg \rho_{0,inj} u_{inj}^t) ,
\end{equation}
and
\begin{equation}\label{ST}
{S_T}_\mu = \partial/\partial t \left(\detg\left(q_{0,inj} u_{inj}^t
u^{inj}_\mu + p_{inj} \delta^t_\mu \right)\right) .
\end{equation}
where $q_{0,inj}\equiv \rho_{0,inj}+u_{inj}+p_{inj}$.  This represents
the source of energy-momentum, such as photon and neutrino losses
emitted and absorbed and the pairs created by annihilation of photons
or neutrinos.  This accounts for, in the guise of the MHD
approximation, ``non-local'' transport of energy and momentum.  The
injected particles are a relativistic gas with $\gamma=4/3$ such that
$p_{inj}=(\gamma-1) u_{inj}$.

Two approximate injection methods are used to treat the injected
momentum.  One assumes that the momentum injected is mostly aligned
with the axis with $u_{inj}^\theta=0$ and either $v_{inj}^\phi=0$ or
$v_{inj}^\phi=\Omega[R=r_{stag}]$ for some typical origin of the
neutrinos $r_{stag}$.  For the other, the particles are injected in
the comoving frame of the existing fluid (i.e. $u_{inj}^\mu=u^\mu$).
It turns out that these different approaches lead to qualitatively
similar results for the solution beyond the region where injection is
important ($r\gtrsim 6r_L$ in the jet).  The total energy density
injected ($\dot{e}_{inj}$ as defined by an observer at infinity) is
partitioned between rest-mass, enthalpy, and momentum energy.  The
fraction of rest-mass injected is defined as $f_\rho$ and so
\begin{equation}
\partial/\partial t (\detg\rho_{0,inj}u^t_{inj}) =
f_\rho\detg\dot{e}_{inj} .
\end{equation}
The fraction of internal energy related injected energy is defined as
$f_h$, and so
\begin{equation}
\partial/\partial t (\detg((u_{inj}+p_{inj})u^t_{inj}
u^{inj}_t+p_{inj})) = f_h\detg\dot{e}_{inj} .
\end{equation}
The fraction of rest-mass plus momentum energy injected is defined as
$f_\rho+f_m$, and so
\begin{equation}
\partial/\partial t (\detg\rho_{0,inj}u^t_{inj}
u^{inj}_t) = (f_\rho+f_m)\detg \dot{e}_{inj} .
\end{equation}
Here $f_\rho+f_h+f_m=1$.

\citet{mckinney2005b} defines $\dot{e}_{inj}$, which along with the
injection fractions and $u^\mu u_\mu =-1$ completely define the
injection process by allowing one to solve for $\rho_{inj}$,
$u_{inj}$, $u^t_{inj}$, and $u_t^{inj}$ for a fixed time interval
$dt$.  Radiative cooling effects are important for studying hundreds
of dynamical times, but otherwise can be neglected if the initial
model is consistent with the time-averaged radiative properties.
Future work will refine the treatment of the pair creation process,
considering the creation effects in the locally flat comoving frame,
once a self-consistent model is available.

\subsection{Pair Plasma Annihilation}

The electron-positron pair plasma that forms may annihilate itself
into a fireball if the pair annihilation rate is faster than the
typical rate of the jet ($c^3/GM$) near the black hole.  Also, if the
pair annihilation timescale is shorter than the dynamical time, then
pair annihilation would give a collisional term in the Boltzmann
equation.  From the pair annihilation rate given in
\citet{mckinney2005b}, one finds that $t_{pa}\gg GM/c^3$ for AGN and
marginally so for x-ray binaries.  Thus, pairs mostly do not
annihilate, and so formally the pair plasma that forms in the
low-density funnel region is collisionless so that the Boltzmann
equation should be solved directly. Plasma instabilities and
relativistic collisionless shocks are implicitly assumed to keep the
pairs in thermal equilibrium so the fluid approximation remains mostly
valid, as is a good approximation for the solar wind (see,
e.g. \citealt{fm97,usmanov2000}).  This same approximation has to be
invoked for the thick disk state in AGN and x-ray binaries, such as
for the ADAF model \citep{mckinney2004}.  For regions that pair
produce slower than the jet dynamical time, each pair-filled fluid
element has a temperature distribution that gives an equation of state
with $P=\rho_{0,e^- e^+} k_b T_e/m_e$ rather than $P=(11/12)a T^4$,
where $a$ is the radiation constant.  So most of the particles have a
Lorentz factor of $\Gamma_e\sim u/(\rho_{0,e^- e^+}c^2)$ and little of
the internal energy injected is put into radiation.  This also allow
the use of a single-component approximation.  A self-consistent
Boltzmann transport solution is left for future work.

On the contrary for GRB systems, due to the relatively high density of
pairs, the time scale for pair annihilation is $t_{pa}\ll GM/c^3$
along the entire length of the jet.  Thus a pair fireball forms and
the appropriate equation of state is that of an
electron-positron-radiation fireball.  Thus, formally the pair
fireball rest-mass density is not independent of the pair fireball
internal energy density.  However, because the pairs are well-coupled
to the radiation until a much larger radius of $r\sim 10^8-10^{10}
r_L$, the radiation provides an inertial drag on the remaining pair
plasma.  That is, the relativistic fluid energy-momentum equation is
still accurate.  So the effective rest-mass density is $\sim \rho_0+u$
($u$ the total internal energy of the fireball), and so the effective
rest-mass is independent of the cooling of the fireball until the
fireball is optically thin (see, e.g., \citealt{mr97}).

For GRB systems, the mass conservation equation is reasonably
accurate.  Even though the electron-positron pairs annihilate, the
rest-mass of pairs injected is approximately that of the pairs that
are injected due to Fick-diffusion of neutrons (see appendix A of
\citealt{mckinney2005b}).  The annihilation energy from
electron-positron pairs contributes a negligible additional amount of
internal energy, so can be neglected, especially compared to the
Poynting energy flux that emerges from the black hole.  Thus, the
rest-mass can always be assumed to be due to baryons rather than the
electron-positron pairs.  This also suggests that the neutrino
annihilation is a negligible effect if the BZ power is larger than the
neutrino annihilation power.

In summary, the rest-mass evolution discussed in section~\ref{results}
is accurate for GRB, AGN, and marginally so for x-ray binaries.  This
is despite the lack of Boltzmann transport for the collisional system,
or a collisional term due to pair annihilation.

\subsection{Electromagnetic Terms}

In terms of $F^{\mu\nu}$, the Faraday (or electromagnetic field)
tensor,
\begin{equation}\label{tmunuem}
T^{\mu\nu}_{\rm EM} =
F^{\mu\gamma}{F^{\nu}}_{\gamma} -\frac{1}{4}g^{\mu\nu}
F^{\alpha\beta}F_{\alpha\beta},
\end{equation}
where a factor of $\sqrt{4\pi}$ is absorbed into the definition of
$F^{\mu\nu}$.  The induction equation is given by the space components
of ${\dF^{\mu\nu}}_{;\nu} = 0$, where $\dF$ is the dual of the
Faraday, and the time component gives the no-monopoles constraint.
The other Maxwell equations, $J^\mu = {F^{\mu\nu}}_{;\nu}$, define the
current density $J^\mu$.  The comoving electric field is defined as
\begin{equation}
e^\nu \equiv u_\mu F^{\mu\nu} = {1\over{2}}\epsilon^{\mu\nu k\lambda} u_\nu \dF_{\lambda k} = \eta j^\nu ,
\end{equation}
where $\eta$ corresponds to a scalar resistivity for a comoving
current density $j^\mu = J_\nu P^{\nu\mu}$, where $P^{\nu\mu}\equiv
g^{\nu\mu} + u^\nu u^\mu$ projects any 4-vector into the comoving
frame (i.e. $P^{\nu\mu} u_\mu = 0$).  The classical ideal MHD
approximation that $\eta = e^\mu=0$ is assumed.  The comoving magnetic
field is defined as
\begin{equation}\label{bcon}
b^\nu \equiv u_\mu \dF^{\mu\nu} = {1\over{2}}\epsilon^{\mu\nu k\lambda} u_\nu \dF_{\lambda k},
\end{equation}
and so the stress-energy tensor can be written as
\begin{equation}
T^{\mu\nu}_{\rm EM} = {b^2\over{2}}(u^\mu u^\nu + P^{\mu\nu}) - b^\mu b^\nu ,
\end{equation}
and
\begin{equation}
\dF^{\mu\nu} = b^\mu u^\nu - b^\nu u^\mu
\end{equation}
and so
\begin{equation}
F^{\mu\nu} = \epsilon^{\mu\nu\sigma\epsilon} u_\sigma b_\epsilon
\end{equation}
since $\dF^{\mu\nu} = {1\over{2}} \epsilon^{\mu\nu\kappa\lambda}
F_{\kappa\lambda}$.  Here $\epsilon$ is the Levi-Civita tensor.
Following the notation of MTW, $\epsilon^{\mu\nu\lambda\delta} =
-{1\over{\detg}} [\mu\nu\lambda\delta]$, where $[\mu\nu\lambda\delta]
$ is the completely antisymmetric symbol and $= 0,1,$ or $-1$.  Notice
that $e^\nu u_\nu = b^\nu u_\nu = 0$, so they each have only 3
independent components and are space-like 4-vectors.

With $B^i \equiv \dF^{it}$ and $E^i \equiv F^{it}$, the no-monopoles
constraint becomes
\begin{equation}
(\detg B^i)_{,i} = 0 ,
\end{equation}
and the magnetic induction equation becomes
\begin{equation}
(\detg B^i)_{,t} = -(\detg(b^i u^j - b^j u^i ))_{,j} .
\end{equation}
The ideal MHD approximation assumes that $e^\mu = 0$, and so the
invariant $e^\mu b_\mu = 0$.  Since the Lorentz acceleration on a
particle is $f_l^\mu=q e^\mu$, then this implies that the Lorentz
force vanishes on a {\it particle} in the ideal MHD approximation.
Equation~\ref{bcon} implies $b^t = B^i u_i$ and $b^i = (B^i + u^i
b^t)/u^t$, so the magnetic induction equation becomes
\begin{eqnarray}
(\detg B^i)_{,t} & = & -(\detg(B^i v^j - B^j v^i)_{,j} \nonumber\\
& = & -(\detg(\epsilon^{ijk}\varepsilon_k))_{,j} ,
\end{eqnarray}
where $v^i=u^i/u^t$, $\varepsilon_i=-\epsilon_{ijk} v^j
B^k=-\bf{v}\times\bf{B}$ is the EMF, and $\epsilon^{ijk}$ is the
spatial permutation tensor.  A more complete account of the
relativistic MHD equations can be found in \cite{anile}.

\section{Characteristic (and other) Surfaces}\label{chars}

The ideal MHD dispersion relation is given, e.g., in \citet{gmt03}
(there is a sign typo there), and summarized here.  In the comoving
frame, the dispersion relation is
\begin{equation}
\begin{array}{l}
D(k^\mu)  = 0 \\
\omega \left(\omega^2 - \kdv^2\right) \times \\
\left(\omega^4 - \omega^2 \,
\left(
K^2 c^2_{ms} + c_s^2 \kdv^2/c^2 \right) + K^2 c_s^2 \kdv^2
\right),
\end{array}
\end{equation}
where $c^2_{ms}=(\va^2 + c_s^2 (1 - \va^2/c^2))$ is the magnetosonic
speed, $c_s^2 = (\del (\rho + u)/\del p)_s^{-1}$ is the relativistic
sound speed, $\bva = \bB/\sqrt{\sE}$ is the relativistic Alfv\'en
velocity, $\sE = b^2 + w$, and $w \equiv \rho + u + p$.  Here $c$ is
the (temporarily reintroduced) speed of light.  The invariant scalars
defining the comoving dispersion relation are $\omega = k_\mu u^\mu$,
$K^2 = K_\mu K^\mu = k_\mu k^\mu+\omega^2$, where $K_\mu = P_{\mu\nu}
k^\nu = k_\mu + \omega u_\mu$ is the projected wave vector normal to
the fluid 4-velocity, $\va^2 = b_\mu b^\mu/\sE$, and $\kdv = k_\mu
b^\mu/\sqrt{\sE}$.  The terms in the dispersion relation correspond
to, respectively from left to right, the zero frequency entropy mode,
the left and right going Alfv\'en modes, and the left and right going
fast and slow modes.  The eighth mode is eliminated by the
no-monopoles constraint.

The dispersion relation gives the ingoing and outgoing slow, Alfv\'en,
and fast surfaces.  Energy can be extracted from the black hole if and
only if the Alfv\'en point lies inside the ergosphere \citep{tak90}.
Optimal acceleration of the flow by conversion of Poynting flux to
kinetic energy flux occurs beyond the outer fast surface
\citep{begli94}.  Other surfaces include: the horizon at $r_H \equiv
1+\sqrt{1-j^2}$ ; the ergosphere at $r\equiv
1+\sqrt{1-(j\cos{\theta})^2}$ ; the coordinate basis light surface in
where $\sqrt{g_{\phi\phi}}=c/\Omega_F$, where asymptotically
$\sqrt{g_{\phi\phi}}=r\sin(\theta)$ is the Minkowski cylindrical
radius ; the surface in Boyer-Lindquist coordinates where the toroidal
field equals the poloidal field where $B^{\hat{\phi}}=B^{\hat{r}}$.
Finally, there is a stagnation surface where the poloidal velocity
$u^p=0$.  In a field confined jet where no matter can cross field
lines into the jet, and if the jet has inflow near the black hole and
outflow far from the black hole, then this necessarily marks at least
one location where rest-mass must be created either by charge starving
the magnetosphere till the Goldreich-Julian charge density is reached,
or pair production rates sustains the rest-mass density.

\end{appendix}

\label{lastpage}

\end{document}